\documentclass[twocolumn,trackchanges]{aastex631}

\usepackage{amsmath,xcolor}
\usepackage{array}

\begin{document}

\title{Irradiated Atmospheres III : \\ Radiative-Convective-Mixing Equilibrium for 
Non-Grey Picket-Fence Model}

\correspondingauthor{Cong Yu}
\email{yucong@mail.sysu.edu.cn}

\correspondingauthor{Dong-dong Ni}
\email{ddni@nju.edu.cn}

\author[0000-0002-0447-7207]{Wei Zhong}
\affiliation{Institute of Science and Technology for Deep Space Exploration, Nanjing University, Suzhou, 215163, People's Republic of China}

\author[0009-0004-1986-2185]{Zhen-Tai Zhang}
\affiliation{School of Physics and Astronomy, Sun Yat-Sen University, Zhuhai, 519082, People's Republic of China}
\affiliation{CSST Science Center for the Guangdong-Hong Kong-Macau Greater Bay Area, Zhuhai, 519082, People's Republic of China}
\affiliation{State Key Laboratory of Lunar and Planetary Sciences, Macau University of Science and Technology, Macau, People's Republic of China}

\author[0000-0002-0378-2023]{Bo Ma}
 \affiliation{School of Physics and Astronomy, Sun Yat-Sen University, Zhuhai, 519082, People's Republic of China}
\affiliation{CSST Science Center for the Guangdong-Hong Kong-Macau Greater Bay Area, Zhuhai, 519082, People's Republic of China}

\author[0000-0003-2278-6932]{Xianyu Tan}
\affiliation{Tsung-Dao Lee Institute \& School of Physics and Astronomy, Shanghai Jiao Tong University, Shanghai 201210, China}

\author[0000-0002-0483-0445]{Dong-dong Ni}
\affiliation{Institute of Science and Technology for Deep Space Exploration, Nanjing University, Suzhou, 215163, People's Republic of China}
\affiliation{State Key Laboratory of Lunar and Planetary Sciences, Macau University of Science and Technology, Macau, People's Republic of China}

\author[0000-0003-0454-7890]{Cong Yu}
\affiliation{School of Physics and Astronomy, Sun Yat-Sen University, Zhuhai, 519082, People's Republic of China}
\affiliation{CSST Science Center for the Guangdong-Hong Kong-Macau Greater Bay Area, Zhuhai, 519082, People's Republic of China}
\affiliation{State Key Laboratory of Lunar and Planetary Sciences, Macau University of Science and Technology, Macau, People's Republic of China}
\affiliation{International Centre of Supernovae, Yunnan Key Laboratory, Kunming 650216, People's Republic of China}

\begin{abstract}

The non-grey picket-fence model predicts more accurately the temperatures in low-density regions compared to semi-grey models. This study investigates how the vertical mixing and convection fluxes modify the picket-fence model. The usual radiative-convective-equilibrium (RCE) is now extended to 
radiative-convective-mixing-equilibrium (RCME). The temperature profile, characterized by an increase with pressure in the upper region and an inversion in the lower, is influenced by Rosseland opacity, spectral bands, and chemical composition.  The atmosphere consists of five distinct layers: a pseudo-adiabatic zone shaped by mixing flux, two convective layers driven by convective flux with a smaller adiabatic gradient, and two radiative layers. In scenarios with lower Rosseland opacity, vertical mixing significantly reduces the width of temperature inversion, counteracting the cooling effect of the convective layers and driving the deep convective layer inward. The convective flux lowers the upper temperature and expands the upper convective layer. In the low-Rosseland-opacity five-band model, these fluxes significantly cool the mid-atmosphere when temperature increases with pressure, enlarging the pseudo-adiabatic region. Without TiO/VO, the pseudo-adiabatic region shrinks, indicating that TiO/VO enhances the mixing effect. Moreover, less mixing intensity is essential to maintain a stable five-layer structure. Therefore, future studies of chemical equilibrium  with multi-frequency atmospheric opacity should clearly define the constraints on vertical mixing.

\end{abstract}

\keywords{{Exoplanet Atmospheres (487) ---Atmospheric structure(2309)---Radiative transfer equation(1336)---Radiative transfer simulations(1967) } }

\section{Introduction}
\label{sec:intro}
More than 5,000 exoplanets have been identified \citep{2022MNRAS.Gupta}, most of which are classified as super-Earths, sub-Neptunes, or hot Jupiters, including ultra-hot Jupiters. These planets exhibit a remarkable diversity in the mass, radius, and orbital distance. The characterization of exoplanetary atmospheres is primarily based on transmission and emission spectrum \citep{2010Seager}. These observations are facilitated by advanced instruments such as the James Webb Space Telescope (JWST; \citealt{2023Carter,2023Miles}) and the Hubble Space Telescope (HST) WFC3 spectrograph, as well as large ground-based facilities. 
These results enable the retrieval of atmospheric properties and improve the accuracy of radiative transfer predictions and analyses.

The vertical mixing causes temperature inversions \citep{Spiegel2009} in ultra-hot Jupiters by keeping Titanium oxide (TiO) and Vanadium oxide (VO) \citep{Hubeny2003ApJ,Fortney2005} from sinking into lower layers, preventing the cold trap formation \citep{Spiegel2009,Parmentier2013,2023Natur.Pelletier,2024arXivZhang}. It reduces methane \citep{2024Natur.630..831S} and somewhat influences cloud \citep{2016AA...Fromang,2018MNRAS...Ryu,2018ApJ...Zhang....1Z,2018ApJ...Zhang,2019MNRAS...Menou,WallaceOrmel2019A&A} and haze formation. While its effects on chemical tracers and mass transport are well-documented, the impact of vertical mixing induced energy transport on radiative transport remains poorly understood. Besides convective zones \citep{Zhang_2020}, vertical mixing also occurs in radiative zones. Gravity waves \citep{Lindzen1981, Strobel1987, Zhang_2020} and large-scale wave-induced mixing \citep{1984maph...Holton, Zhang_2020} aid chemical tracer mixing in upper layers. Investigating energy transport by vertical mixing in radiative layers \citep{Youdin2010} is vital for enhancing atmospheric models and understanding energy dynamics on exoplanets.

Convection significantly impacts the atmosphere. Large-scale convection \citep{Zhang_2020} is crucial for vertical mixing. Convection also modifies heat distribution by changing the temperature gradient in the convective layer \citep{Robinson2012Ap}. The convective layer can also reduce excess temperature increases from lower levels \citep{Wallace1977}. Moist convection \citep{Seeley2023PSJ} may arise from atmospheric chemistry but can be destabilized by convective inhibition \citep{Cavali2017Icar,Leconte2024A&A}. Therefore, understanding the role of convection is vital for the radiative atmosphere. Unlike convective temperature adjustments based directly on the adiabatic gradient \citep{Robinson2012Ap,Parmentier2015, ZhangXi2023ApJ}, the modification of convective temperature via convective flux \citep{Gandhi2017} from mixing-length theory is more reasonable, as this approach must match radiative equilibrium to conserve energy. Thus, their interaction can change atmospheric structure.

For the purpose of simplifying radiative transfer analyses, semi-grey models \citep{2010A&A...Guillot} are frequently utilized, which segment the spectrum into thermal and optical bands. Nonetheless, these models tend to overestimate temperatures in atmospheric layers characterized by low density \citep{Parmentier2015}.
Non-grey models, like the picket-fence model \citep{Chandrasekhar1935}, excel by simulating a wider temperature range and providing more accurate predictions of temperature at low optical depth than semi-grey models \citep{2014A&A...Parmentier,Parmentier2015}. 
In summary, the improved accuracy and broader applicability of the picket-fence model render the combined effects of vertical mixing and convective fluxes essential for atmospheric research.

This study examines the combined effects of vertical-mixing and convective fluxes on the non-grey picket-fence atmospheric model.
Atmospheric circulation \citep{Zhang_2020} or the breaking of gravity waves \citep{Strobel1987, Zhang_2020} drive vertical mixing, creating downward flux \citep{Youdin2010}. These fluxes establish a radiative-convective-mixing equilibrium (RCME). In this model, the temperature initially rises with pressure, then inverts, forming two convective layers. The reduced Rosseland opacity under RCME enhances these signatures. Vertical mixing reduces the width of the temperature inversion via a greenhouse-like effect \citep{Zhong2024}, while convective flux cools the lower atmosphere and alters the upper convective layer. Adding more visible bands and chemical elements improves the temperature accuracy of the model. According to \citet{Spiegel2009,2024Natur.630..831S,Welbanks2024}, vertical mixing intensity ranges between \(10^6\) and \(10^{11}\ \mathrm{cm^2\ s^{-1}}\). As the temperature is sensitive to chemical composition changes, our model could help constrain the vertical mixing intensity.

The structure of this study is articulated as follows. The function of the RCME in conjunction with the radiative transfer equation, as affected by the interaction between vertical mixing induced energy transport and convective flux, is demonstrated in \S~\Ref{Picket_Fence_Model}. Moreover, the effects of vertical mixing and convective flux on the configuration of the non-grey picket-fence atmosphere are analyzed in \S~\Ref{sec_resluts}. Ultimately, the conclusions and discussions are detailed in \S~\ref{sec_conclusion}.

\section{Picket-Fence Model} \label{Picket_Fence_Model}

An irradiated atmosphere is a planet's atmosphere affected by external radiation, particularly from a stellar source. The fluxes from both vertical mixing in the radiative and convective layers can lead to a RCME, altering the atmospheric structure. Radiative transfer in the picket-fence model is detailed in \S~\ref{sec_RTE} and the RCME is in \S~\ref{SEC_RCME}. The simulation method is in \S~\ref{sec:simulation}.

\subsection{Radiative Transfer for the picket-fence model}\label{sec_RTE}
In this section, we briefly describe the picket-fence model, which is distinct from the semi-grey atmosphere. 
Generally, the radiative transfer equation \citep{2014tsa..book...Hubeny,Parmentier2014PhDT,Gandhi2017} satisfies 
\begin{equation}
     \frac{d H_{\nu}}{d\tau_{\nu}} = \kappa_{\nu}-B_{\nu}\  ,
     \label{eq:dHv_dtauv}
\end{equation} and 
\begin{equation}
     \frac{d K_{\nu}}{d\tau_{\nu}} = H_{\nu}\ .
     \label{eq:dKv_dtauv}
\end{equation}
Here, $J_{\nu}$, $H_{\nu}$, and $K_{\nu}$ represent the zeroth, first, and second moment intensity at different wavelength frequencies $\nu$, respectively. Moreover, $\kappa_{\nu} $ denotes the opacity at a specific wavelength. Additionally, $B_{\nu}$ is the Planck function for the temperature $T$, and $\tau_{\nu}$ signifies the optical depth at various frequencies. The atmosphere is assumed to be in radiative equilibrium \citep{2010A&A...Guillot,2014A&A...Parmentier,2014tsa..book...Hubeny}, as illustrated by 
\begin{equation}
     \int_{0}^{\infty} \kappa_{\nu} \left( J_{\nu}-B_{\nu}\right) = 0 \ .
     \label{eq_blance}
\end{equation}

To facilitate simplification, it is imperative to employ mean opacities. 
The Rosseland mean \citep{2014A&A...Parmentier,2014tsa..book...Hubeny}, acknowledged as the most commonly used, is formulated as
\begin{equation}
    \kappa_{\rm R} = \int_{0}^{\infty }\frac{\partial B_{\nu}}{\partial T} d\nu \left(\int_{0}^{\infty } \frac{1}{\kappa_{\nu}}\frac{\partial B_{\nu}}{\partial T} d\nu\right)^{-1} \ .
\end{equation}
In circumstances where the mean free path of photons is considerably less than the atmospheric scale height, the radiative gradient adheres to the diffusion limit. 
Consequently, this results in the alignment of the temperature gradient with that of a grey atmosphere utilizing the Rosseland mean opacity.
Therefore, optical depth is vertically characterized utilizing the Rosseland mean in the following manner \citep{2014A&A...Parmentier}:
\begin{equation}
    d\tau \equiv \kappa_{\rm R} dm = - \kappa_{\rm R} \rho dz\ , 
    \label{eq_tau}
\end{equation}
with column mass $m$, density $\rho$, and the altitude $z$. 
Under the assumption of hydrostatic equilibrium, we should integrate Equation (\ref{eq_tau}) to ascertain the relationship between pressure and optical depth:
\begin{equation}
    \tau\left(P\right) \equiv  \int_{0}^{P} {\kappa_{\rm R} \left(P^{\prime
    }, T^{\prime}\right)}/{g} d P^{\prime} \ ,
\end{equation}
with constant gravity $g$ and pressure $P$. The optical depth emerges as the natural variable for representing depth dependence in the radiative transfer problem. In addition, the second mean opacity commonly applied in radiative transfer theory is referred to as the Planck mean \citep{2014A&A...Parmentier,2017MNRAS...Hubeny}:
\begin{equation}
    \kappa_{\rm P} \equiv  \left(\int_{0}^{\infty} B_{\nu} d\nu \right)^{-1} \int_{0}^{\infty} \kappa_{\nu} B_{\nu} d\nu \ .
\end{equation}
The ratio of Planck to Rosseland mean opacities measures the atmosphere's non-greyness, i.e.,
\begin{equation}
    \gamma_{\rm P} = \kappa_{\rm P}/\kappa_{\rm R} \ .
    \label{eq_gamma_p}
\end{equation}
While the Rosseland mean opacity is primarily governed by the lowest values of the opacity function $\kappa_{\nu}$, the Planck mean opacity is largely influenced by its highest values. 
Thus, $\gamma_{\rm P} = 1$ applies to a grey atmosphere, and $\gamma_{\rm P}>1$ applies to a non-grey atmosphere. The ratio of visible opacity $\kappa_{\rm v}$ to the Rosseland mean thermal opacity is articulated as follows:
\begin{equation}
    \gamma_{\rm v} = \kappa_{\rm v}/\kappa_{\rm R} \ .
\end{equation}
To solve the radiative transfer problem analytically, assume $\gamma_{\rm v}$ is constant with optical depth. With $\gamma_{\rm v}$ set, visible radiation equations are solvable independently of the thermal structure. In purely grey models, $\gamma_{\rm v}=1$ applies.

The non-grey picket-fence model offers a comprehensive temperature profile and accurately predicts temperature distribution at low optical depths. It is applicable in scenarios that account for both upward and downward radiation, rendering it highly relevant for temperature fitting and prediction.
Following \citet{2014A&A...Parmentier}, the picket-fence model incorporates three distinct opacities: two different values $\kappa_{\rm 1}$ and $\kappa_{\rm 2}$ for the thermal radiation (the thermal opacities), and $\kappa_{\rm v}$ for incoming radiation from the star. 
$\kappa_{\rm 1}$ and $\kappa_{\rm 2}$ are associated with an equivalent bandwidth $\beta$. 
In reference to the Rosseland opacity, these ratios are specified as $\gamma_{\rm 1} = \kappa_{\rm 1}/\kappa_{\rm R}$ and $\gamma_{\rm 2} = \kappa_{\rm 2}/\kappa_{\rm R}$. Furthermore, a relationship pertinent to thermal radiation is delineated by $R = \gamma_{\rm 1}/\gamma_{\rm 2} = \kappa_{\rm 1}/\kappa_{\rm 2}$.
Note that the quantities $R$, $\gamma_1$, and $\gamma_2$ are depth independent.

Two different sets of opacities, each characterized by varying wavelength dependencies, may yield equivalent Rosseland and Planck mean opacities.
The Planck and Rosseland mean opacities \citep{2014A&A...Parmentier} are expressed as
\begin{equation}
    \kappa_{\rm P} = {\beta \kappa_{\rm 1} + \left(1-\beta\right)\kappa_{\rm 2}} \ ,
    \label{eq:10}
\end{equation}
and
\begin{equation}
    \kappa_{\rm R} = \frac{\kappa_{\rm 1}\kappa_{\rm 2}}{\beta \kappa_{\rm 2} + \left(1-\beta\right)\kappa_{\rm 1}} \ ,
        \label{eq:11}
\end{equation}
respectively.
The model recognizes three types of opacity: low and high picket-fence opacity, and "visible" opacity. Note that, when defining the expressions for the Rosseland and Planck mean opacities (i.e., Equations. \ref{eq:10} and \ref{eq:11}), we only consider the relevant averaged opacity in the thermal band. The "visible" opacity is also taken into account, which determines the values of $\gamma_{\rm v}$ and the intensity signature for the visible band according to Equation (\ref{eq:dKv_dtau}).
Changes in \( R \) and \( \beta \) directly affect the temperature profile, as shown by the shaded area in Figure~\ref{fig_Model_validation}. Here, \( R \) and \( \beta \) are treated as independent variables, influencing other related parameters. A detailed explanation is provided in \S~\ref{sec:validation}.
Moreover, the Equations~(\ref{eq:dHv_dtauv}-\ref{eq:dKv_dtauv}) for the radiative transfer equations will be transformed into 
\begin{equation}
    \frac{d H_{\rm 1}}{d\tau} = \gamma_{\rm 1}\left(J_{\rm 1} - \beta B  \right) \ ,
    \label{eq:dH1_dtau}
\end{equation}
\begin{equation}
    \frac{d H_{\rm 2}}{d\tau} = \gamma_{\rm 2}\left[J_{\rm 2} - \left(1-\beta \right) B \right]\ ,
    \label{eq:dH2_dtau}
\end{equation}
\begin{equation}
    \frac{d H_{\rm v}}{d\tau} = \gamma_{\rm v}J_{\rm v} \ ,
    \label{eq:dH2_dtau1}
\end{equation}
\begin{equation}
    \frac{d K_{\rm 1}}{d\tau} = \gamma_{\rm 1} H_{\rm 1} \ ,
    \label{eq:dK1_dtau}
\end{equation}
\begin{equation}
    \frac{d K_{\rm 2}}{d\tau} = \gamma_{\rm 2} H_{\rm 2}\ ,
    \label{eq:dK2_dtau_new}
\end{equation}
\begin{equation}
    \frac{d K_{\rm v}}{d\tau} = \gamma_{\rm v}H_{\rm v}\ ,
    \label{eq:dKv_dtau}
\end{equation}
in which the parameters denoted by the subscripts ``$1$'', ``$2$'', and ``${\rm v}$'' pertain to the momentum associated with the dual-band thermal radiation and the incident stellar radiation, respectively.
In addition, $B$ represents the mean Planck function derived via integration across thermal wavelengths.
An equation governing the conservation of total flux is represented in \cite{2010A&A...Guillot}, i.e.,
\begin{equation}
    H = \int_{0}^{\infty} H_{\nu} d\nu = H_{\rm 1} + H_{\rm 2} + H_{\rm v} = {\rm const } \ .
    \label{eq_flux_conservation}
\end{equation}
Coupled with Equations~(\ref{eq:dH1_dtau})-(\ref{eq:dH2_dtau1}), this previously mentioned equation is reformulated into \begin{equation}
    \gamma_{\rm 1} J_{\rm 1} + \gamma_{\rm 2} J_{\rm 2} + \gamma_{\rm v} J_{\rm v}  = \gamma_{\rm P}B \ ,
    \label{eq_flux_conservation_2}
\end{equation}
which is referred to as the radiative equilibrium equation. 
The Eddington approximation, $J_{\left(1,2\right)} = 3 K_{\left(1,2\right)}$, is a standard closure relationship. 
For the top boundary ($\tau \sim 0$), the intensity of zero radiation moment is  set as 
\begin{equation}
    J_{\left(1,2\right)} \left(0\right) = 2 H_{\left(1,2\right)}\ .
    \label{eq_top_boundary_01}
\end{equation}
Equation (\ref{eq_top_boundary_01}) defines the boundary conditions for the two thermal bands. However, the picket-fence model includes incoming stellar irradiation. Our boundary conditions incorporate this effect by modifying the outer boundary for the optical bands. Specifically, stellar irradiation alters the boundary condition for the optical band as follows: $ H_{\rm v}\left(0\right) = -\sigma_{\rm R} T_{\rm irr}^4/4\pi $ \citep{2010A&A...Guillot,2014A&A...Parmentier}, where $ T_{\rm irr}$ represents the effective temperature of irradiation and $\sigma_{\rm R} $ is the Boltzmann constant. For multiple optical bands, the effect of stellar irradiation is distributed proportionally among the $n$ bands, each with width $\beta_{\rm vi}$, as described in Equation (\ref{eq:40}).  
At the bottom boundary \citep{2014A&A...Parmentier}, the condition is 
\begin{equation}
    \frac{\partial B}{\partial \tau} \sim 3 H_{\infty} = 3\sigma_{\rm R} T_{\rm int}^4/4\pi \ ,
    \label{eq_bottom_boundary_01}
\end{equation}
with the intrinsic temperature $T_{\rm int}$.

\subsection{The Radiative-Convective-Mixing Equilibrium}\label{SEC_RCME}

The interplay between convection and vertical mixing in a planetary atmosphere is pivotal for determining its atmospheric structure. Previous researches on radiative transfer have largely overlooked the integrated effects of vertical mixing within the radiative zone and the flux from the convection layer.
The resultant flux from mixing and convection modifies the radiative equilibrium. In the radiative zone, where convective flux is absent, mixing alters the radiative equilibrium (RE), leading to a state referred to as the radiative-mixing equilibrium (RME). In contrast, within the convection layer, the mixing flux decreases, resulting in the radiative-convective equilibrium (RCE). To clarify, we refer to their combined effect across the entire envelope as the radiative-convective-mixing equilibrium (RCME).

The flux of vertical mixing can trigger an RME within the radiative zone. In the absence of external factors, heat typically transfers from hotter to cooler regions, in accordance with the second law of thermodynamics. However, near the infrared (IR) photosphere, vertical mixing is driven either by atmospheric circulation or by the breaking of gravity waves. 
\citet{Youdin2010} investigated the effects of the additional heat flux produced by the mixing. This phenomenon induces entropy mixing, resulting in heat transfer from colder to warmer regions and subsequently increasing atmospheric temperatures \citep{Youdin2010,2018Leconte}. The expression for the mixing flux is given by
\begin{equation}\label{eq.F}
    F_{\rm mix}=-K_{\rm zz}\rho g\left(1-\frac{\nabla}{\nabla_{\rm ad}}\right) \ ,
\end{equation}
with a gravitational acceleration fraction represented by $g=2500 \  cm \ s^{-2}$. The vertical mixing strength is denoted by \( K_{\rm zz} \).
Moreover, the logarithmic temperature gradient is defined as 
\begin{equation}\label{nabla}
    \nabla= \frac{d\ln T }{d\ln P}=\frac{P}{T}\frac{dT}{dP} \ .
\end{equation}
The adiabatic gradient applicable to an ideal diatomic gas is represented by $\nabla_{\rm ad} = 2/7$.
The value of  $K_{\rm zz}$, is influenced by atmospheric properties. 
The RME instigated by vertical mixing fluxes is articulated as
  \begin{equation}
     H_{\rm 1}+H_{\rm 2}+H_{\rm v}+\frac{F_{\rm mix}}{4\pi}=\frac{\sigma_{\rm R}}{4\pi}T^4_{\rm int} \ ,
     \label{eq_blance}
 \end{equation}
 \begin{equation}
     \gamma_{\rm 1} J_{\rm 1} + \gamma_{\rm 2} J_{\rm 2} + \gamma_{\rm v} J_{\rm v}  - \gamma_{\rm P}B +\frac{1}{\kappa_{\rm R}}\frac{\rho g}{4 \pi}\frac{d F_{\rm mix}}{dP}=0 \ .
     \label{nbalance2}
 \end{equation}
Following \citet{Gandhi2017}, we apply Equation (\ref{eq_blance}) for the lower atmosphere and Equation (\ref{nbalance2}) for the upper atmosphere to enhance the numerical stability of our calculation. These equilibrium equations are discussed in $\nabla<\nabla_{\rm ad}$.

Convection occurs in atmospheres where the Schwarzschild criterion is met $\nabla$ $>$  $\nabla_{\rm ad}$. 
It is vital as it often exceeds radiative transfer in carrying energy through deeper layers. High optical depth and low radiative flux reduce the efficiency of energy transport, making it essential to identify convective zones and calculate the flux for model integration.

We adopt the mixing length theory (MLT) \citep{2013book..Kippenhahn} to describe convection. As a gas parcel rises, it rapidly reaches thermal equilibrium with its surroundings, leading to adiabatic cooling. If the radiative temperature gradient exceeds the adiabatic gradient, the parcel remains warmer and continues to ascend, indicating convective instability. This process alters the RE, forming a RCE, where conditions Equations~(\ref{eq_flux_conservation}) and (\ref{eq_flux_conservation_2}) include extra flux from convection,
  \begin{equation}
     H_{\rm 1}+H_{\rm 2}+H_{\rm v}+\frac{F_{\rm conv}}{4\pi}=\frac{\sigma_{\rm R}}{4\pi}T^4_{\rm int} \ ,
     \label{eq_blance4}
 \end{equation}
 \begin{equation}
     \gamma_{\rm 1} J_{\rm 1} + \gamma_{\rm 2} J_{\rm 2} + \gamma_{\rm v} J_{\rm v}  - \gamma_{\rm P}B +\frac{1}{\kappa_{\rm R}} \frac{ \rho g}{4 \pi}\frac{d F_{\rm conv}}{dP}=0 \ .
     \label{nbalance3}
 \end{equation}
Here, the convective flux $F_{\rm conv}$, as defined by mixing length theory \citep{2013book..Kippenhahn}, is expressed as
\begin{equation}
\begin{aligned}
   F_{\rm conv} 
   \left(\nabla, T, P\right) =  
   \ F_{\rm 0} \left(\nabla-\nabla_{\rm el}\right)^{3/2} \ ,
\end{aligned}
\end{equation}
where
\begin{equation}
    F_0 = \left(\frac{g Q H_{\rm P}}{32}\right)^{1/2} \rho c_{\rm P} \left(l/H_{\rm P}\right)^2 \ , 
\end{equation}
\begin{equation}
    \nabla-\nabla_{\rm el} = \frac{A^2}{2}  + \xi - A \left(\frac{A^2}{4}+\xi\right)^{1/2}\ .
\end{equation}
Note that in the above equation, 
\begin{equation}
    A \equiv \frac{16 \sqrt{2}\sigma_{\rm R}T^3}{\rho c_{\rm P}\left( g Q H_{\rm  P}\right)^{1/2}\left(l/H_{\rm P}\right)} \frac{\tau_{\rm el}}{1+\frac{1}{2}\tau_{\rm el}^2} \ ,
\end{equation}
and $\xi = \nabla - \nabla_{\rm ad}$.
In this context, $c_{\rm P}$ denotes the heat capacity at constant pressure, and $Q \equiv -\left(d\ln \rho/d \ln T\right)_{\rm P}$ represents a parameter that equals 1 for an ideal gas. The optical depth $\tau_{\rm el} = l \kappa_{\rm R}$ of a diminutive gas parcel with dimension $l$ is considered here as well, where $l$ is a variable often approximated to be $\approx H_{\rm P}$, indicating the atmospheric scale height. The specific choice of $l$ exerts minimal influence on the observed flux of hot Jupiters, due to the position of convective zones typically being well beneath the visible atmosphere. The elemental logarithmic temperature gradient, denoted as $\nabla_{\rm el}$, fulfills the condition
\begin{equation}
    \nabla_{\rm el}-\nabla_{\rm ad} = A \sqrt{\nabla -\nabla_{\rm el} }\ . 
\end{equation}
By adding $\nabla$ to both sides and rearranging, $\nabla -\nabla_{\rm el}$ can be calculated from the resultant quadratic in $\sqrt{\nabla -\nabla_{\rm el}}$, thus
\begin{equation}
    \nabla-\nabla_{\rm ad} = \nabla -\nabla_{\rm el} + A\sqrt{\nabla -\nabla_{\rm el}} \ .
\end{equation}

\subsection{Numerical  Method}\label{sec:simulation}

For the efficient determination of temperature profiles, the second order radiative transfer Equations (\ref{eq:dH1_dtau})-(\ref{eq:dKv_dtau}) can be expressed as follows:
\citep{Gandhi2017,2017MNRAS...Hubeny}
\begin{equation}
 \frac{\partial^2 K_{\nu}}{\partial \tau_\nu^2} = \frac{\partial^2\left(f_\nu J_{\nu}\right)}{\partial \tau_\nu^2} 
 = \gamma_{\nu}^2 \left(J_{\nu}-B_{\nu}\right) \ ,
     \label{eq:RTE}
\end{equation}
using the Eddington factor $f_\nu = K_\nu/J_\nu = 1/3$ for thermal and $f_\nu = K_\nu/J_\nu = \mu^2$ for visible bands.
$\mu$ is the cosine of the irradiated incident angle, which affects the second Eddington factors $f_\nu$ and $g_\nu$.
The notation $\nu$ for frequency bands is consistent with \S \ref{sec_RTE}, which refers to bands $1$, $2$, and ${\rm v}$. 
In addition, $B_{\rm 1} = \beta B$, $B_{\rm 2} = \left(1-\beta\right)B$, and $B_{\rm v}=0$.
In our current model, for $B_{\rm v} = 0$, visible radiation is assumed to be absorbed without emission for simplicity. However, Rayleigh scattering can be a significant or even dominant source of opacity in the visible range. The omission of scattering is a limitation of our approach and may impact its applicability. Neglecting the visible scattering can result in lower temperatures in the upper atmosphere and higher temperatures in the lower atmosphere than when scattering is included.  The role of scattering is reported elsewhere\footnote{The interplay between scattering and vertical mixing in irradiated atmospheres, Zhang et al. (2025), submitted.}.
The Equation~(\ref{eq_top_boundary_01}) for the top boundary of the atmosphere is transformed into 
\begin{equation}
     \left.\frac{\partial\left(f_{\nu} J_{\nu}\right)}{\partial \tau_{\nu}}\right|_{\tau_\nu=0}= \gamma_{\nu}g_\nu J_{\nu 0}-\gamma_{\nu}H_{\mathrm{ext}}\ .
     \label{eq:upper_boundary}
\end{equation}
The second Eddington factor $g_{\nu}$ is set to $1/2$ for the thermal band and $-\mu$ for the visible band.
Equation~(\ref{eq_bottom_boundary_01}) for the bottom boundary of the atmosphere changes into
\begin{equation}
     \left.\frac{\partial\left(f_{\nu} J_{\nu}\right)}{\partial \tau_{\nu}}\right|_{\tau_{\max }}=\gamma_{\nu} \left[\frac{1}{2}\left(B_{\nu}-J_{\nu}\right)+\frac{1}{3} \frac{\partial B_{\nu}}{\partial \tau_\nu}\right]_{\tau_{\max}} \ .
\label{eq:bottom_boundary}
\end{equation}
Density is calculated via the ideal gas law $P = {\rho k_{\mathrm{b}} T}/{\overline{m}}$, incorporating both the average molecular mass $\overline{m}$ and the Boltzmann constant $k_{\mathrm{b}}$. 
The radiative transfer equations become non-linear when adherent to the RME condition. 
To address the nonlinear equations (\ref{eq:RTE})-(\ref{eq:bottom_boundary}) and (\ref{nbalance2})-(\ref{eq_blance4}), Rybicki's technique \citep{2014tsa..book...Hubeny,2017MNRAS...Hubeny,2019PhDT...Gandhi} is employed. 
The Rybicki scheme is a method used for temperature correction in RCE/RCME models, employing a complete linearization approach to ensure accurate convergence to the correct atmospheric temperature profile. It iteratively adjusts the temperature until the RCE/RCME condition is satisfied, improving computational efficiency and stability in exoplanetary atmosphere modeling.  For more details, see \S 3.3 in \citet{2017MNRAS...Hubeny} or \S 17.3 in \citet{2014tsa..book...Hubeny}. In this work, we find that the results obtained using the Rybicki scheme and the traditional complete linearization method \citep{1969ApJAuer...Mihalas} are identical. However, we will implement the Rybicki scheme in order to assess its performance in handling cases with a very large number of frequencies (tens or hundreds of thousands) in the future, particularly to study the effects of chemistry in high-resolution atmospheric models.

\section{Results} \label{sec_resluts}

The study explores the effects of the RCME on the picket-fence model of planetary atmospheres. \S~\ref{sec:validation} focuses on validating the model without convection and mixing, by comparing simulation results with analytical solutions from \citet{2014A&A...Parmentier}. \S~\ref{sec:VCE} investigates changes in atmospheric structure due to vertical mixing and convection. \S~\ref{sec:kappa_r} analyzes the impact of the Rosseland opacity on temperature profiles and then gains an understanding of RCME. Finally, we assess the five-band picket-fence model with and without TiO/VO in \S~\ref{SEC_FIVE_BAND}.

\begin{figure}
    \centering
    \includegraphics[width= 1.1\linewidth]{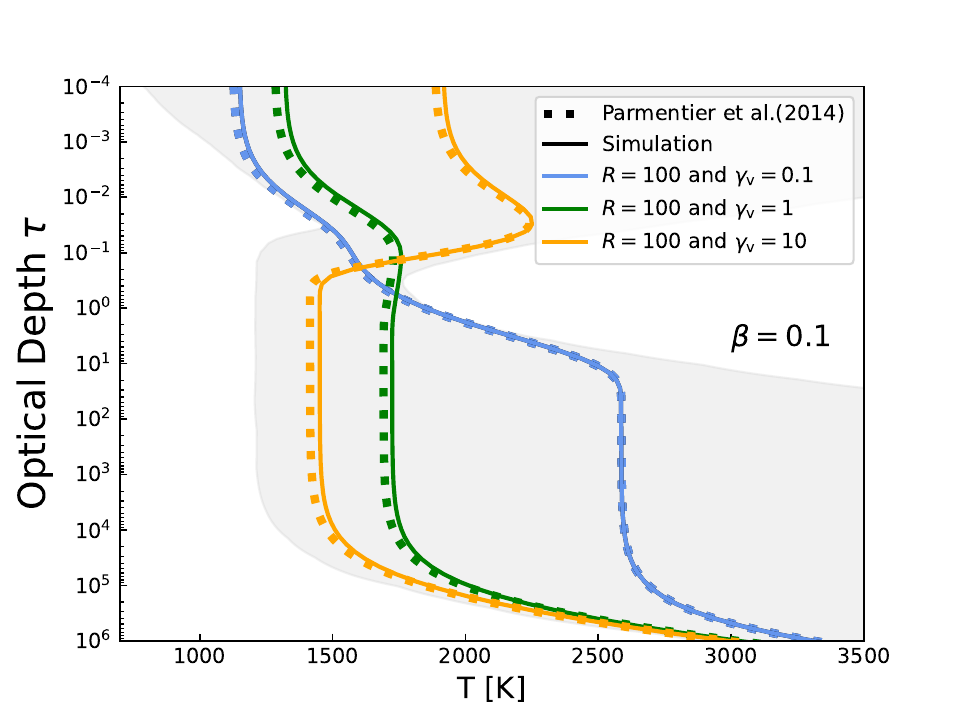}
    \caption{A temperature-optical depth diagram without vertical mixing and convection is shown for $\mu = 1/\sqrt{3}$. Dotted lines represent the analytical solutions from Equation (76) of \citet{2014A&A...Parmentier}, while solid lines represent numerical simulation results of this study. The shaded area indicates the full range of parameters $10^{-3} < \beta < 10^{-1}$, $1 < R < 10^4$, $0.01 < \gamma_{\rm v} < 100$. Profiles for $\beta = 0.1$ appear as plain lines for $R = 100$, blue for $\gamma_{\rm v} = 0.1$, green for $\gamma_{\rm v} = 1$, and orange for $\gamma_{\rm v}$ = 10.
    }
    \label{fig_Model_validation}
\end{figure}

\subsection{Validating Numerical Treatment}\label{sec:validation}

We validate our calculation of the picket-fence model by comparing with the analytical solutions from Equation (76) of \citet{2014A&A...Parmentier}. 
The results are shown in Figure~\ref{fig_Model_validation}.
Using \( R \) and \( \beta \) as separate variables, we modify the analytical model, following \cite{2014A&A...Parmentier}\footnote{The original model is accessible at https://cdsarc.u-strasbg.fr/viz-bin/qcat?J/A+A/574/A35}. \citet{2014A&A...Parmentier} noted that non-grey atmospheres exhibit lower temperatures at reduced optical depths compared to semi-grey atmospheres. As \( \beta \) approaches one, deeper layers experience warming induced by a blanketing effect. By setting \( R = 10^2 \) and \( \beta = 0.1 \), we derive gamma parameters that are consistent with the analytical expression, aligning with the Eddington approximation and the boundary conditions established in both models.

The visible gamma parameter $\gamma_{\rm v} = \kappa_{\rm v}/\kappa_{\rm R}$ affects the temperature profile by dictating opacity at the visible band $\kappa_{\rm v}$. Lower visible opacity leads to a monotonic  temperature increase with pressure, as radiation penetrates easily.
However, increased visible opacity can cause a mid-atmosphere temperature inversion, unlike the monotonic temperature increase or complete inversion in the semi-grey model. 
All picket-fence model cases fall within the shaded area, and solutions converge at high optical depths irrespective of non-grey opacity variations. 
The close match between the simulation results and the analytical solutions suggests that our simulation of the picket-fence model is correct.

\subsection{Effect of Vertical Mixing and Convection} 
\label{sec:VCE}

\begin{figure*}
    \centering
    \includegraphics[width=1.0\linewidth]{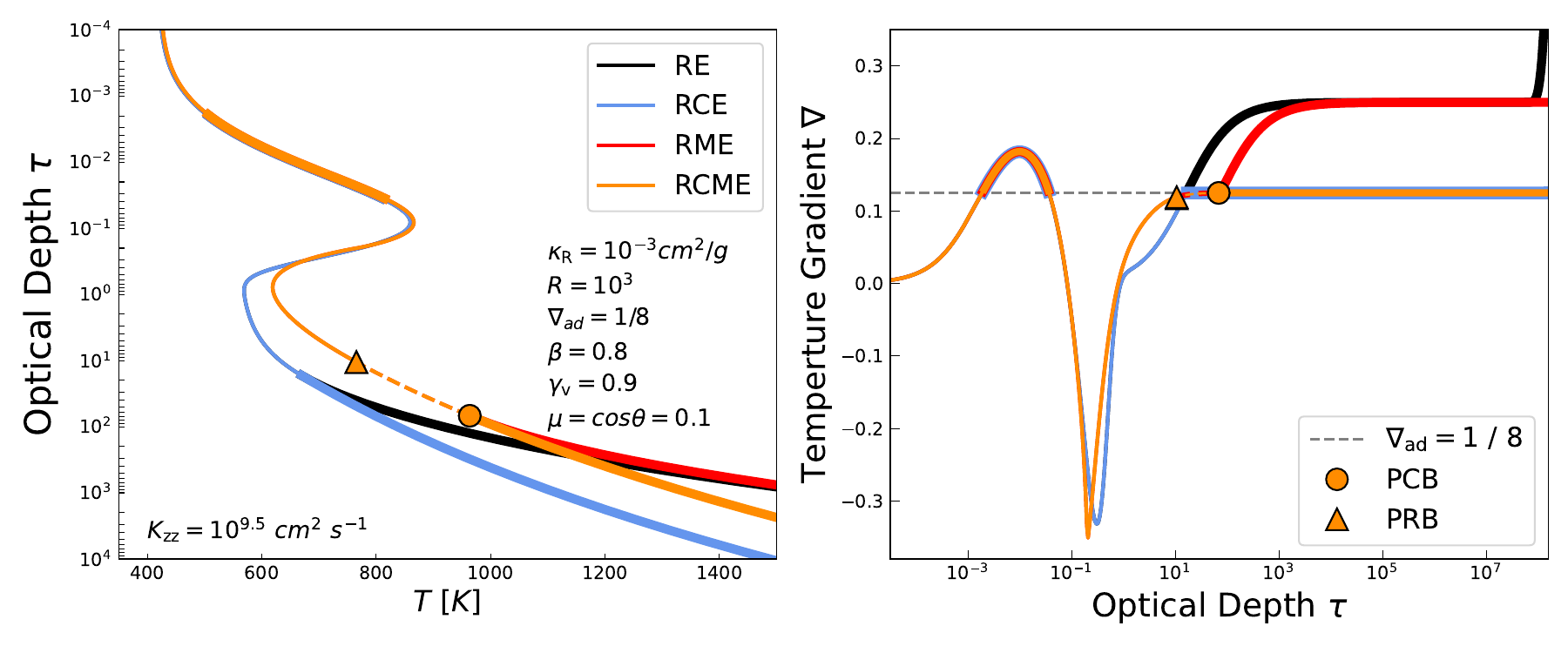}
    \caption{
The left panel shows the temperature-depth profiles for various equilibrium states.
The right panel illustrates the relationship between the temperature gradient and optical depth for each case.
Each curve represents a different atmospheric condition: black for radiative equilibrium (RE), blue for radiative-convective equilibrium (RCE), red for radiative-mixing equilibrium (RME), and orange for radiative-convective-mixing equilibrium (RCME).
In both panels, narrow lines indicate the radiative region, while thicker lines represent the convective region. For the case without convective flux, the convection instability region is defined by where convection becomes unstable, without considering the temperature modification.
Dashed lines mark the pseudo-adiabatic region. Triangular markers indicate the boundary between the pseudo-adiabatic and radiative regions (PRB), while circular markers represent the boundary between the pseudo-adiabatic and convective regions (PCB).
    }
    \label{fig_Pseudo_rcb}
\end{figure*}

This section examines how convection and vertical mixing alter initial temperature profiles in a picket-fence model, where the initial profile is divided into regions exhibiting rising temperatures and a temperature inversion zone. 
Utilizing the methodology outlined in \S~\ref{sec:simulation}, we investigate the combined effects of vertical mixing and convection on a RE atmosphere. 
In our simulation, we adopt the gamma coefficient ratio of the thermal bands as \( R = 10^3 \), effective bandwidth as \( \beta = 0.8 \), Rosseland opacity as \( \kappa_{\rm R} = 10^{-3} \ cm^2/g \), and the ratio of visible to Rosseland opacity at \( \gamma_{\rm v} = 0.9 \).
Moreover, the irradiated temperature, internal temperature, and cosine of the incident angle are fixed at $T_{\rm irr}= 1000$K, $T_{\rm int}=300$ K, and $\mu=cos \theta=0.1$, respectively.
We show the simulated temperature profile and its corresponding gradient in Figure~\ref{fig_Pseudo_rcb}. 
Adding vertical mixing and convection causes significant changes in both the temperature profile and their gradients.

Convection-induced flux is vital in the deeper atmosphere. Rather than modifying the temperature profile of the convective layer using the adiabatic gradient \(\nabla_{\rm ad}\) \citep{Parmentier2015} directly, we incorporate convective fluxes into a RE framework \citep{2014tsa..book...Hubeny, Gandhi2017, 2017MNRAS...Hubeny}. Utilizing the Schwarzschild criterion, we categorize the atmosphere into radiative (stable) and convective (unstable) layers. In convective regions, \citet{Parmentier2015} modified the adiabatic gradient with the high-pressure equation from \citet{Saumon1995}, approximating it as dry adiabatic \(\nabla_{\mathrm{ad}} \approx 0.32 - 0.1(T/3000 \, \mathrm{K})\). It was observed that reducing this gradient  \(\nabla_{\rm ad}\) significantly affects temperature, as illustrated in Figures 5–7 of \citet{Parmentier2015}.
In the right panel of Figure~\ref{fig_Pseudo_rcb}, the temperature gradient of the RE atmosphere (black line) increases almost linearly near the inner boundary, which exceeds the ideal dry adiabatic gradient of 2/7.
To delineate the effects of RCE, we assume a fixed convection adiabatic temperature gradient \(\nabla_{\rm ad} = 1/8\).
Within our model shown in Figure~\ref{fig_Pseudo_rcb}, the narrow lines are indicative of the radiative zone, whereas the thick solid lines denote the convective zone.

Unlike the single convective layer with a monotonic temperature rise with pressure in \cite{Parmentier2015}, Figure~\ref{fig_Pseudo_rcb} displays two convective layers linked by a smaller \(\nabla_{\rm ad}\) in this particular temperature profile.
In the absence of convective energy, we just apply the Schwarzschild criterion to delineate these zones without introducing any temperature adjustments.  
In this configuration, the black and red lines depict scenarios where the temperature gradient exceeds \(\nabla_{\rm ad}\): one around \(\tau \sim 10^{-3} - 10^{-1}\) and another at \(\tau > 10^{1}\) or $10^{2}$. Although only part of the black line is visible in Figure~\ref{fig_Pseudo_rcb}, its trend aligns with the blue line in the left panel. 
The thick blue line in the left panel highlights a sharp temperature drop in the inner convective zone, driven by a reduced gradient. 
In contrast, the outer convective layer stays stable, with minimal convective flux ensuring consistent temperature (left panel) and gradient (right panel).
However, a lower $\kappa_{\rm R}$ can alter the temperature and gradient of this zone, as will be discussed in \S~\ref{sec:kappa_r}.

The picket-fence atmosphere can also be influenced by the induced flux resulting from the vertical mixing within the radiative layer  \footnote{The impact of vertical mixing-induced flux on the semi-grey atmosphere is demonstrated in \citet{Zhong2024}.}. 
The vertical mixing strength within the atmosphere is estimated to span a broad range, from $10^6$ to $10^{11.6} \, {\rm cm}^2 \, {\rm s}^{-1}$. This estimate consolidates findings from earlier studies and recent models based on observational constraints \citep{Spiegel2009,Sing2024Natur,Welbanks2024}. Our tests show that, with a larger \(T_{\rm int}\) and smaller \(T_{\rm irr}\), a stronger mixing intensity \( K_{\rm zz} = 10^{9.5} \, {\rm cm}^2 \, {\rm s}^{-1} \) is required in the picket-fence model to clearly manifest the mechanical greenhouse effect.
The results are represented by the red curve in Figure~\ref{fig_Pseudo_rcb}.
This mechanical greenhouse effect raises temperatures in the lower atmosphere (as shown by the red line compared to the black line) and forms a pseudo-adiabatic region, where $\nabla$ closely approaches $\nabla_{\rm ad}$.
However, the upper atmosphere still remains stable.
Therefore, the width of the temperature inversion is reduced.
Despite these changes, the temperature at the bottom boundary remains consistent with non-mixing models, resulting in a five-layer atmospheric structure. 
The pseudo-adiabatic and radiative boundaries (PRB) are marked by triangles, while pseudo-adiabatic and convective boundaries (PCB) are marked by circles. 
Vertical mixing also shifts the innermost convective layer inward.

The introduction of convective energy modifies the radiative-mixing equilibrium (i.e., RME), leading to a new equilibrium that incorporates radiation, convection, and mixing (i.e.,  RCME). 
This adjustment alters the planetary atmospheric temperature profile. 
As shown in the right panel, 
the temperature gradient of the convective layer aligns with the adiabatic gradient, with the orange line overlapping the grey dashed line. 
Correspondingly, the temperature decreases significantly in the left panel. 
However, the convective energy in the upper atmosphere is insufficient to impact its temperature, thus the outermost convective layer remains consistent with other scenarios. 
In comparison to the case with RCE, the lower atmosphere experiences additional heating.

\begin{figure*}
    \centering
    \includegraphics[width=1.0\linewidth]{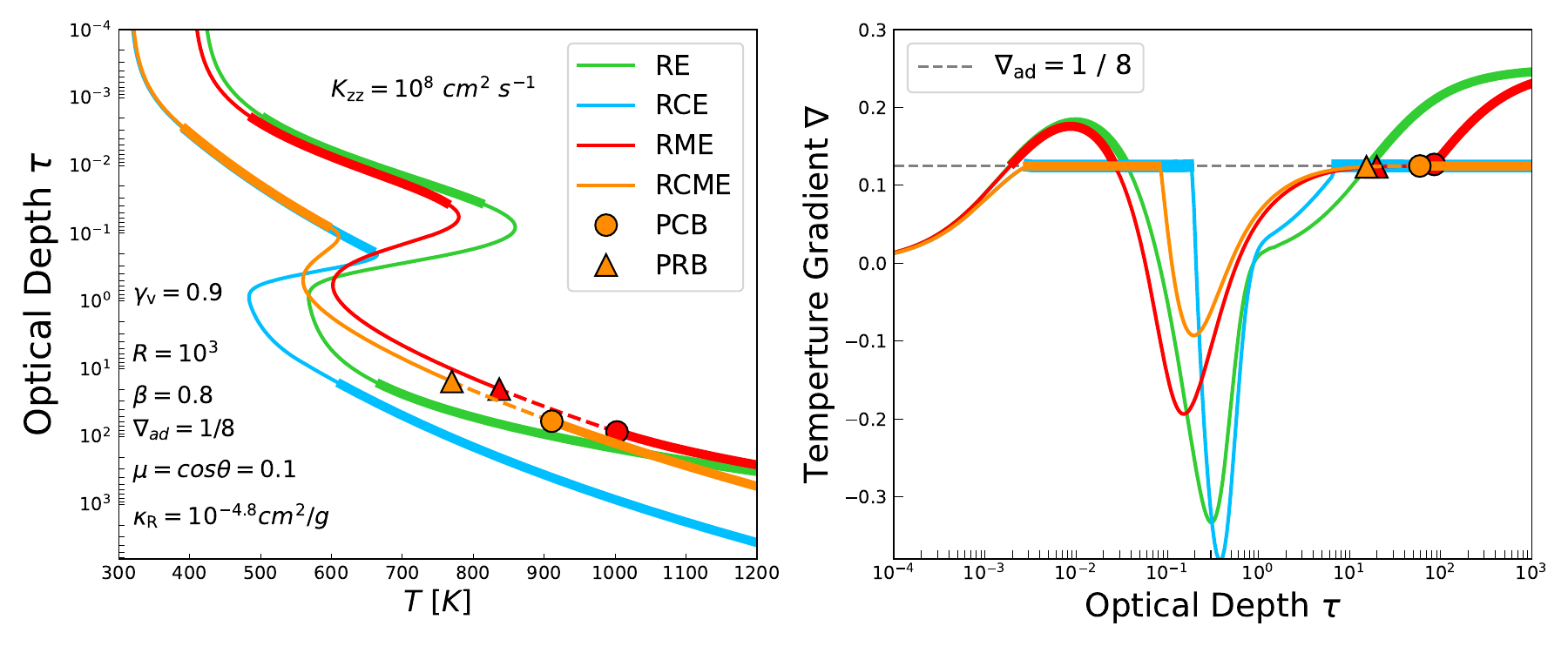}
    \caption{Reduced Rosseland opacity influences the RCME temperature structure. 
    In the absence of convection and mixing, radiative equilibrium (RE) is reached, as illustrated by the lime-green solid line. 
    When convection is included, RCE is portrayed by the light-blue line. 
    Vertical mixing transforms it to RME, shown by the red line. 
    Together, convection and mixing form RCME, represented by the orange line. 
    The boundary between pseudo-adiabatic and radiative regions is triangular; it becomes circular between pseudo-adiabatic and convective regions. 
    The meanings of used acronyms is the same as in Figure \ref{fig_Pseudo_rcb}.}
    \label{fig:Low Rosseland opacity}
\end{figure*}

Our simulation demonstrates that the interaction between vertical mixing energy and convective energy plays a significant role in shaping the planetary structure.
Traditionally, mixing intensity is evaluated using spectral analysis through model retrieval, often overlooking the effect of vertical mixing induced energy transport. 
Since there is a need to maintain energy balance when considering the impact of vertical mixing induced energy transport, we can put a more accurate constraint on the mixing intensity from the planetary temperature profiles. 
Our model shows that a high mixing intensity significantly heats the lower atmosphere, while the convective flux reduces the temperature in the innermost convective region. The interaction between vertical mixing and convective energy plays a crucial role in shaping the planetary structure.

\subsection{Effect of Rosseland Opacity}\label{sec:kappa_r}

In this section, we examine the impact of low Rosseland opacity on the structure of planetary atmospheres. 
We show the simulation results in Figure~\ref{fig:Low Rosseland opacity}.
Without additional energy input, the temperature gradient of the RE atmosphere (right panel) shows two zones (i.e., the convective layer) that surpasses the adiabatic gradient, similar to Figure~\ref{fig_Pseudo_rcb}. 
The left panel shows two convective regions marked with thick lime-green lines, which will vary when convection and vertical mixing are considered.

\begin{figure}
    \centering
    \includegraphics[width= 1.1\linewidth]{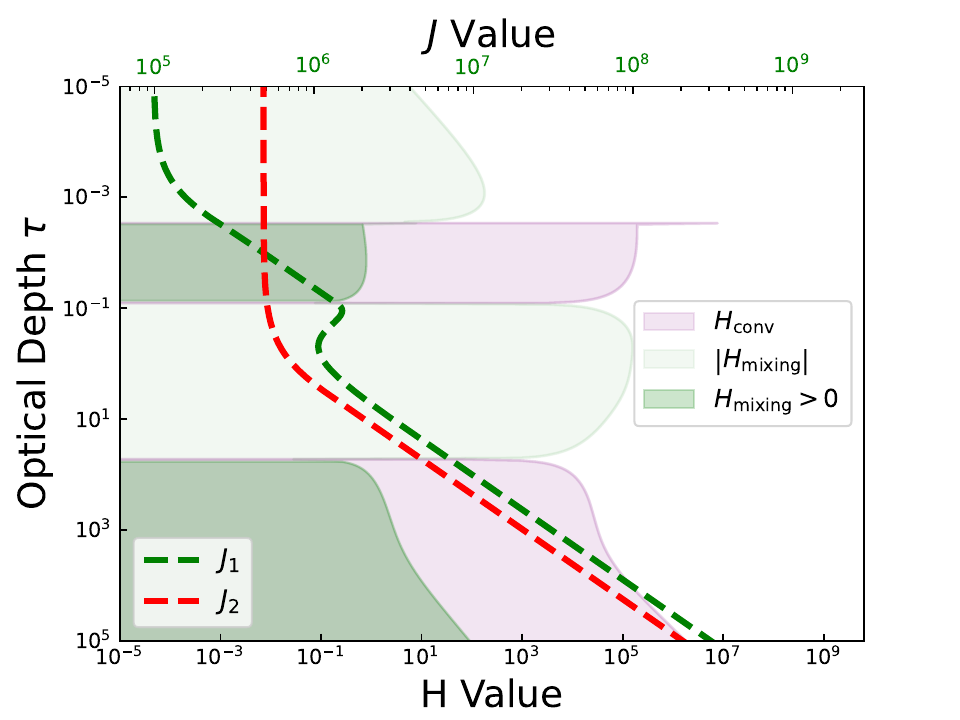}
    \caption{Profiles of convective and mixing energy optical depths (bottom X-axis) are shown with slope maps of thermal band radiation intensification (top X-axis). Light green indicates regions with negative vertical mixing energy, while dark green denotes positive vertical mixing energy. The light purple area represents convective energy, which is positive. The green dotted line outlines the radiation intensity of the thermal band 1, and the red line depicts that of thermal band 2.
    }
    \label{fig:kappaR with energy}
\end{figure}

In atmospheres with low Rosseland opacity, the convective flux modifies the entire temperature profile, especially in regions where $\nabla > \nabla_{\rm ad}$, as indicated by the thick lime-green line for the RE case in the right panel.
The strength of convective energy is influenced by the Rosseland opacity. As shown in the RCE Equation (\ref{nbalance3}), a lower opacity leads to stronger convective energy. Our tests show that when $\kappa_{\rm R} \le  10^{-4.8}\ \text{cm}^2/\text{g}$, the increased convective energy effectively corrects the temperature gradients in both convective regions.
The convection energy  contour ($H_{\rm conv}>0$) is akin to the light purple area in Figure~\ref{fig:kappaR with energy} for the RCME atmosphere. 
Convective energy at greater depths adjusts the temperature gradient of the convective layer (the thick lime-green line) to match the adiabatic gradient.

The convective energy flux notably impacts the upper atmosphere in this section.
Figure~\ref{fig:Low Rosseland opacity} illustrates the outward shift of the innermost radiative-convective boundary (RCB, where the thick and thin lines intersect), leading to the divergence of the temperature for the light-blue line from the lime-green line. 
As we move upward to lower depth, this outward shift decreases the temperature in the inversion region compared to the RE scenario shown by the lime-green line.
The following convective flux occurs near $\tau \sim 10^{-1}-10^{0}$ as the temperature gradient steepens. 
Compared to the upper regions where $\nabla > \nabla_{\rm ad}$ in the non-mixing case, the convective flux broadens the convective region and shifts the RCBs inward in the mid-atmosphere.
The temperature gradient in this region approaches $\nabla_{\rm ad}$, with a noticeable temperature drop observed in the left panel at this optical depth. As a result of these changes, the temperature in the upper atmosphere decreases compared to the case of  RE atmosphere(the lime-green line).

The mixing flux is absent here.
Thus, the outer temperature adheres to the RCE Equation (\ref{nbalance3}) at the top boundary, i.e., the change in \(\ J_{\left(1,2\right)}\left(0\right) = 2 H_{\left(1,2\right)}\left(0\right) \)\ .
Based on Equation~(79.21) in \citet{Mihalas..1984} and derivations by \citet{2014A&A...Parmentier}, the top flux boundary conditions for these bands are:
\begin{equation}
       J_{\left(1,2\right)}\left(0\right) = \int_0^{\infty} S_{\left(1,2\right)} E_{\rm 2}\left(\tau \right) d\tau \ ,
\label{eq_anlysis_boundary}
\end{equation}
where the source functions are $S_{\rm 1}\left(\tau \right)= \beta B\left(\tau/\gamma_{\rm 1}\right)$ and $S_{\rm 2} = \left(1-\beta\right)B\left(\tau/\gamma_{\rm 2} \right)$, assuming local thermodynamic equilibrium. 
Moreover, $E_{\rm 2}\left(x\right)$ refers to the second exponential integral within the series
\begin{equation}
E_{\rm n} \left(x\right) \equiv \int_{0}^{\infty} y^{-n}e^{-xy}dy = x^{n-1}\int_{x}^{\infty} y^{-n}e^{-y}dy\ .
\end{equation}
In the upper atmosphere, $E_{\rm n}(x) \sim 1$. With $\beta = 0.8$, temperature fluctuations are caused by changes in $J_{1}(0)$, which result from variations in $S_{\rm 1}$ within the modified uppermost convection layer.
Introducing convective flux into the low-Rosseland-opacity RE atmosphere reduces $J_{1}(0)$ compared to scenarios without convection. As a result, the boundary temperature decreases due to energy conservation, leading to a much lower temperature in the upper atmosphere, where $T \sim T_{\rm out}$.

The orange line in Figure~\ref{fig:Low Rosseland opacity} highlights how vertical mixing modifies the RCE temperature profiles.
To accurately assess the influence of vertical mixing on an RCE atmosphere, it is essential to initially ascertain its effects on an RE (the lime-green line) atmosphere.
Similar to the increase in convective energy under low Rosseland opacity, introducing the same vertical mixing intensity also leads to a stronger vertical mixing flux in the low-Rosseland-opacity atmospheres, as described by the RME Equation~(\ref{nbalance2}). To better illustrate the effects of vertical mixing, we selected a lower mixing intensity of $K_{\rm zz} = 10^8\ cm^2\ s^{-1}$ for this simulation, compared to the mixing intensity used in Figure~\ref{fig_Pseudo_rcb}.

At $\tau \sim 0.5$, the mixing flux gradient $\left|d F_{\rm mix}/d\tau \right|$ for the red line reaches its maximum. Vertical mixing raises temperatures in the lower atmosphere, creating pseudo-adiabatic zones and resulting in new five-layer structures.
The upper boundary of the temperature inversion will push outward.
Additionally, the innermost convection zones are expected to form at greater depths. At higher altitudes, minor temperature decreases are observed, particularly where the lime-green and red lines converge.

Convective flux in the RME atmosphere is expected to cause significant temperature deviations compared to predictions from other models.
The temperature profile is more strongly affected by $J_{\rm 1}$  in Figure~\ref{fig:kappaR with energy} as the term $\gamma_2 (1 - \beta) J_{\rm 2}$ in the energy flux conservation equation becomes smaller.
Consistent with prior analysis, the negative mixing flux in the light green zone of the upper atmosphere is not strong enough to impact $J_{\rm 1}$.
Consequently, the top temperature of the RCME atmosphere (orange line) in Figure~\ref{fig:Low Rosseland opacity} matches that of the RCE atmosphere (light blue line).

In the deeper unstable convective layer, the positive $H_{\rm mixing}$ is counteracted by the convective flux, leading to a cooling effect compared to the RME atmosphere.
When $J_{\left(1,2\right)}$ reaches $\tau \sim 0.5$ in the second negative $H_{\rm mixing}$, the bottom atmosphere starts to rise. Compared to the RME atmosphere, the positions of PRB and PCB in the RCME atmosphere are displaced outward due to a decrease in $H_{\rm mixing}$ caused by convective flux.
As a result, the middle radiative layer cools, which affects the width of the outermost convective layer. Although its width is smaller than in the RCE atmosphere, it remains larger than in the RE atmosphere.

In the RCME atmosphere, the cooling of the upper layers is largely due to convective flux, while vertical mixing results in warming within the lower layers. 
We propose that the interaction between convection and mixing is more significant in atmospheres with lower Rosseland opacity.

\subsection{The five-band picket-fence model}
\label{SEC_FIVE_BAND}

Our model begins with a simple single visible band, but it can be generalized to $n$ visible bands.
The linearity of Equations  (\ref{eq:dHv_dtauv}) and (\ref{eq:dKv_dtauv}) in the visible spectrum allows for solutions across any combination of bands \citep{2014A&A...Parmentier,Parmentier2015}.
The visible intensity momentum is given by:
\begin{equation}
J_{\mathrm{v}}(\tau)=-\frac{H_{\mathrm{v}}(0)}{\mu_*} \sum_{{\rm i}=1}^n \beta_{\mathrm{vi} } \mathrm{e}^{-\gamma_{\rm v i}^* \tau}\ ,
\end{equation}
where $\beta_{\mathrm{vi}}$ is the width of the $i$th band, $\kappa_{\mathrm{vi} }$ is the opacity, and $\gamma_{\rm v i}^* =\gamma_{\rm vi}/\mu$. 
Constant opacities are chosen to approximate the absorption of stellar flux at various heights.
The absorbed flux, derived from $n$ bands of width $\beta_{\rm vi}$, is in agreement with the flux obtained through numerical simulation, as shown below
\begin{equation}
  F(\tau)=F_0 \sum_{\rm i=1}^n \beta_{\rm vi} \mathrm{e}^{-\gamma_{\mathrm{vi}} \tau / \mu_*}  \ .
  \label{eq:40}
\end{equation}
Here, $F_0$ denotes the total incoming flux, and $\beta_{\rm vi}$ must satisfy the relationship $\sum_{\rm i=1}^n \beta_{\rm vi}$. 
In Figure 8 of \citet{Parmentier2015}, it is clear that a single band does not sufficiently capture the absorbed flux. The use of two, three, and four bands achieves precisions of 4\%, 1\%, and 0.5\%, respectively. Hence, we opt for three bands of equal width, $\beta_{\mathrm{vi}} = 1 / 3$.
The choice of gamma parameters has a significant effect on the convective flux and vertical mixing in the five-band picket fence model. In \S~\ref{subsec_constant_gamma}, we examine their effects when constant, and in \S~\ref{subsec_chemistry}, we explore the impact when chemistry is included.

\subsubsection{The Atmosphere with Constant gamma Parameters}
\label{subsec_constant_gamma}

\begin{figure}
    \includegraphics[width=1.1\linewidth]{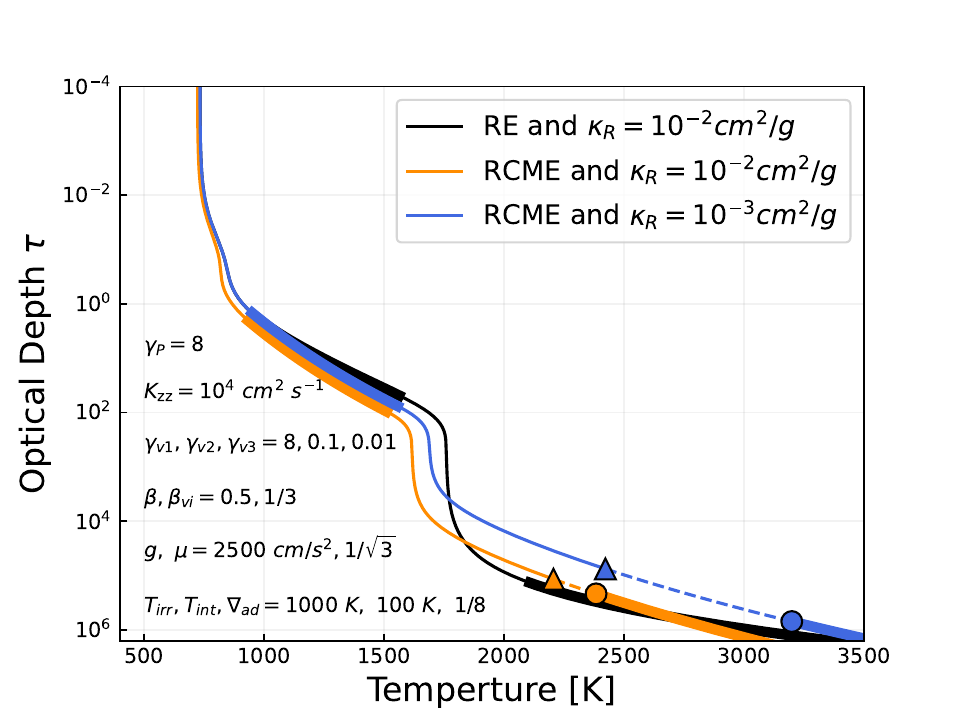}
    \caption{ The influence of vertical mixing and convection in a five-band picket-fence model is simulated using constant gamma parameters across various $\kappa_{\rm R}$ values.
    The radiative region and convective layer are represented by narrow and thick lines, respectively. 
    The pseudo-adiabatic region is indicated by the dashed line. 
    PRB and PCB are denoted by Triangle and Circle, respectively. 
    The black solid line depicts the atmosphere without vertical mixing and convection when $\kappa_{\rm R}=10^{-2} \ cm^2/g$.
    The orange line shows changes in signatures due to vertical mixing and convection, while the blue line depicts atmospheric results with $\kappa_{\rm R}=10^{-3} \ cm^2/g$.
    }
    \label{fig:five_band_constant}
\end{figure}

We investigate the impact of increasing the visible band on the temperature profile by employing constant gamma parameters in a five-band picket-fence model.
The gamma parameters for the three visible bands are $\gamma_{\rm vi} = 8,\ 0.1,\ 0.01$, with $\beta = 0.5$, $\beta_{\rm vi} = 1/3$, $\gamma_{\rm P} = 8$ for the Planck opacity, the constant Rosseland opacity $\kappa_{\rm R} =10^{-2}\ cm^2/g$,  and a vertical mixing intensity of $K_{\rm zz} = 10^4\ \text{cm}^2 \ \text{s}^{-1}$. Other parameters are set as shown Figure~\ref{fig_Pseudo_rcb}.
The final result is shown in Figure~\ref{fig:five_band_constant}.

The combined impact of mixing and convective flux on the temperature is represented by the orange curves, with the black line showing the case where neither convection nor mixing occurs. 
Similar to \S~\ref{sec:VCE}, this non-convective model exclusively utilizes the Schwarzschild criterion for the demarcation of zones.
In these circumstances, convective energy dominates the upper atmosphere, while the lower atmosphere is shaped by the interaction of mixing and convective flux.
Robust convective energy forces the innermost RCB of RE atmosphere outward, which cools the radiative zone at lower depth, as illustrated by the change from the thin black curve to the orange curve within the range $\tau \sim 10^2-10^{4}$.
In parallel, vertical mixing produces a mechanical greenhouse effect, which warms the deeper atmosphere and forms a pseudo-adiabatic region. This causes the convective layer to move to greater depths, as indicated by the thick orange curve.
Similar to \S~\ref{sec:kappa_r}, the convective energy in this region is strong enough to completely adjust it to the adiabatic limit. The outermost convective layer, represented by the thick orange line, will extend inward.
The cooling effect from convection narrows the outermost convective layer, resulting in the outward movement of the RCB.

The lower mean Rosseland opacity (\(\kappa_{\rm R}\)) influences the combined effect of vertical mixing and convective flux on the structure of atmospheric temperature.
RCME proposes that as the effect of vertical mixing increases, the impact of convective energy decreases.
When \(\kappa_{\rm R}\) is high, the impact of vertical mixing on the temperature increase near the bottom of the atmosphere will reduce.
However, a decrease in \(\kappa_{\rm R}\)  enhances the effect of vertical mixing, resulting in a higher base temperature, a larger pseudo-adiabatic zone, and a deeper convective zone compared to the condition shown by the orange line.
The weakening of convective energy correction slows the temperature decrease at the base, while the convective zone at the upper atmosphere (illustrated by the thick blue line) slightly extends.

\subsubsection{The Influence of Gamma Parameters with the Chemistry on Atmosphere}
\label{subsec_chemistry}

The albedo, thermal, and visible coefficients vary with chemistry.
Model D of \citet{Parmentier2015} accurately forecasts deep atmospheric temperatures and establishes coefficients using the effective temperature.
Here, the effective temperature is defined as:
\begin{equation}
T_{\mathrm{eff}}^4=T_{\mu_*}^4+T_{\mathrm{int}}^4 \ ,
\end{equation}
with intrinsic temperature $T_{\rm int}$. 
The temperature \(T_{\mu_*}\) is derived from an isotropic approximation \citep{2010A&A...Guillot, Parmentier2015}, as shown in \begin{equation}
T_{\mu_*}^4=\left(1-A_{\mu_*}\right) 4 f T_{\mathrm{eq0}}^4 \ . 
\end{equation}
$f$ is less than one. For $f=0.25$, the thermal profile aligns with the average of the planet. For $f=0.5$, it matches the day-side average.
For a planet with zero albedo, the equilibrium temperature is given by:
\begin{equation}
T_{\mathrm{eq0}}^4 \equiv \frac{T_*^4}{4}\left(\frac{R_*}{a}\right)^2 \ ,
\end{equation} 
where $T_*$ is the effective temperature of the star, $R_*$ is its radius, and 
 $a$ is the distance from the star.
The plane-parallel albedo $\log_{10}\left(A_{\mu_*}\right)=a+b X$ is determined by $X=\log_{10}\left(T_{\text {eff0 }}\right)$ and $g$.
Albedo depends on Rayleigh scattering and absorption, varying with titanium/vanadium oxides (TiO/VO) and sodium/potassium chemistry.
Albedo can indirectly indicate atmospheric scattering effects.
The values of $a$ and $b$ change with the range \(T_{\mathrm{eq0}}\), as shown in Table 2 of \citet{Parmentier2015}.

The coefficient functions $\gamma_{\rm v1}$, $\gamma_{\rm v2}$, $\gamma_{\rm v3}$, and $\gamma_{\rm P}$, along with the width parameter $\beta$, were provided in Table 1 of \citet{Parmentier2015} for an atmospheric model with solar composition. 
The values of $\gamma_{\rm 1}$ and $\gamma_{\rm 2}$ can also be derived from Equations~(92-93) and (96) of \citet{2014A&A...Parmentier}.
On irradiated planets, the high temperatures on the day side enable 
TiO/VO to stay stable in the gas phase of a solar composition atmosphere \citep{Lodders2002ApJ}.
However, vertical \citep{Spiegel2009} and horizontal cold traps \citep{Parmentier2013}, stellar radiation decomposition \citep{Knutson2010ApJ}, or high C/O ratios \citep{Madhusudhan2012ApJ} can deplete TiO and VO in the upper atmosphere \citep{Parmentier2014PhDT}.
The presence of TiO and VO greatly influences atmospheric radiation absorption. Without these compounds, the values of  \(\gamma_{\rm v1}\), \(\gamma_{\rm v2}\) and \(\gamma_{\rm v3}\) decrease 10 times,  which causes energy to be absorbed more deeply.

\begin{figure}
    \includegraphics[width=1.1\linewidth]{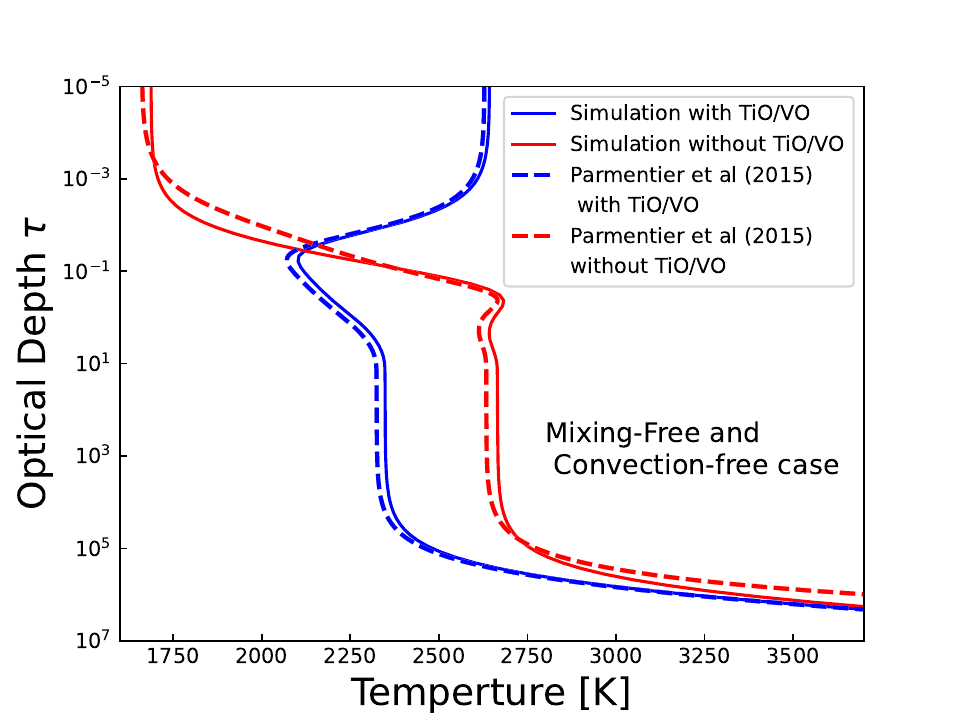}
    \caption{ 
The impact of chemical elements on the temperature profile in the absence of convective and vertical mixing processes. The solid line represents the result obtained from the analytical solution as referenced in \citet{Parmentier2015}, while the dotted line illustrates the outcome of our numerical simulation. The blue line denotes the scenario involving TiO/VO, whereas the red line indicates the scenario without these compounds.
    }
    \label{fig_five_band_2}
\end{figure}

The accuracy of our model was validated by comparing it with the analytical results of \citet{Parmentier2015}, as shown in Figure~\ref{fig_five_band_2}.
We adapt the parameters from their FORTRAN code \footnote{A FORTRAN code of the analytical model from \citet{Parmentier2015} available online at http://cdsarc.u-strasbg.fr/viz-bin/qcat?J/A+A/574/A35.}.
This section assumes the internal temperature $T_{\rm int} = 100\ K$, the equilibrium temperature at zero albedo $T_{\rm eq0}  = 2268\ K$, the planet-averaged profile $f = 0.25 $ and gravity $g = 2500\ cm/s^2$, with no convection layer taken into account.
We employed a constant $\kappa_{\rm R} = 10^{-2}\ cm^2/s$ to simulate the nonlinear radiative transfer equations. 
The resultant temperature profile closely resembles that calculated with the $\kappa_{\rm R}$ interpolated from \citet{Valencia2013ApJ}.

As seen in Figure~\ref{fig_five_band_2}, in the absence of TiO/VO, the solid red line, which represents the analytical results, shows a monotonic rise in temperature with pressure, with a small inversion in the mid-layer.
The red dashed line from the numerical simulation closely matches the analytical results, with only a minor discrepancy. 
The solid blue analytical line separates from the solid red line in the presence of TiO/VO.
These molecules warm the upper atmosphere and cool the deeper layers by absorbing stellar radiation and reducing the thermal shielding effect.
This leads to a thermal inversion, caused by the increased stellar radiation absorption and reduced cooling efficiency in the deeper atmosphere.
The close match between the blue dashed line in our simulations and the analytical results validates the correctness of our model.

\begin{figure}
    \includegraphics[width=1.1\linewidth]{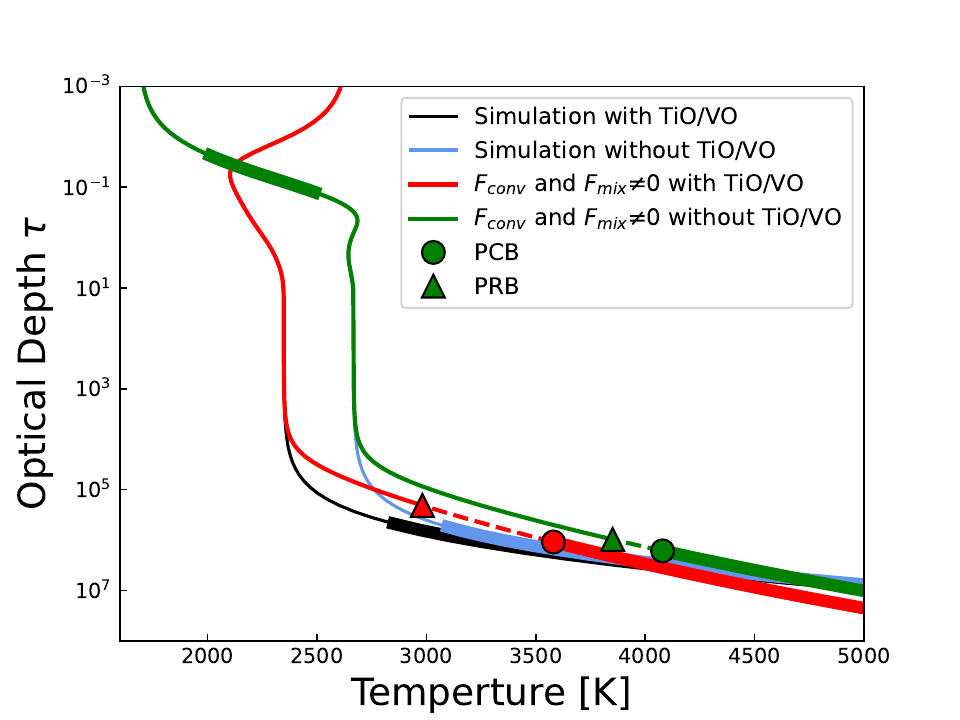}
    \caption{
The impact of vertical mixing and convective flux on atmospheres with or without TiO/VO. Black/red lines show TiO/VO atmospheres with/without these effects, while blue/green lines show TiO/VO-free atmospheres with/without these effects. 
Thin and thick lines indicate radiative and convective layers, respectively. The dashed line shows the pseudo-adiabatic region. Triangles and circles indicate the PCB and PRB, overlapping with their positions in atmospheres with vertical mixing.
    }
\end{figure}

Vertical mixing and convective processes determine temperature profiles, whether or not TiO/VO are involved.
The black line represents the non-mixing and non-convection case, where convective zones are determined solely by the Schwarzschild criterion without any temperature modifications.
In addition, the black temperature inversion profile shows that the upper atmosphere is not limited by any constraints on convection because the temperature gradient is maintained below \(\nabla_{\rm ad} = 1/9\).
However, the temperature in the lower atmosphere can be restricted by convective flux.
To better show the temperature changes in different layers of the atmosphere, the vertical mixing intensity is set at $K_{\rm zz} = 10^2\ cm^2/s$.
In the same way as in the previous section, vertical mixing initiates the mechanical greenhouse effect, elevating the temperature of the lower atmosphere and establishing a pseudo-adiabatic layer between the radiative and convective zones.
With the addition of convective energy, the temperature of the innermost convective zone is further reduced. As a result, while the inversion region becomes narrower, the bottom temperature decreases.

In the absence of TiO/VO, a convective layer forms in the upper atmosphere, but despite the presence of convective flux, the temperature of the radiative zone remains unchanged. The overlapping green and blue lines highlight this result.
We expect that a reduction in $\kappa_{\rm R}$ will affect the temperature in the upper convection layer, potentially altering the overall results, as shown in  Figure~\ref{fig:kappaR with energy}.
The effect of vertical mixing flux diminishes in the absence of TiO/VO, resulting in a more restricted pseudo-adiabatic region, as indicated by the green dashed line, compared to the wider regions of the red lines. This demonstrates that TiO/VO plays a critical role in enhancing vertical mixing, particularly in the deeper atmosphere.

Vertical mixing strength, denoted as $K_{\rm zz}$, is crucial in exoplanet atmospheres. It's mainly due to mass exchange through convection or turbulence at rates of around $10^6-10^{11.6}\ cm^2\ s^{-1}$. 
The role of $K_{\rm zz}$ in energy transfer within the radiative zone, through mechanisms like gravity waves or circulation, is not well understood.
$K_{\rm zz}$ might be lower in certain areas than in convective zones.
Semi-grey models tend to overestimate high atmospheric temperatures, while non-grey models don't.
A smaller $K_{\rm zz}$ is enough for temperature profiles in non-grey models with 3 or 5 bands.
Figure~\ref{fig:kappaR with energy} shows how chemistry influences the atmospheric structure. 
To make the lower convective zone visible, we've set a lower $K_{\rm zz}$. 
This indicates that the mixing strength is influenced by the chemical composition.
To improve predictions of $K_{\rm zz}$, we plan to update our radiation transfer code. We'll include chemical equilibrium and disequilibrium to better match empirical observations.

\section{Conclusion and discussion}\label{sec_conclusion}

The non-grey picket-fence model provides more precise and lower upper atmospheric temperatures than the semi-grey model. We introduce convection and vertical mixing to RE, creating RCME. This research investigates if RCME changes the temperature profiles in three-band and five-band picket-fence models. The picket-fence model shows a distinct temperature profile: temperatures increase monotonically with pressure in the upper atmosphere, and a temperature inversion occurs in the lower atmosphere. The combined effect of vertical mixing and convection on the temperature profile is shaped by frequency bands, $\kappa_{\rm R}$, and chemical composition.

Unlike the approach in \citet{Parmentier2015}, which uses the adiabatic gradient to directly alter convective temperatures, our model integrates convective flux into radiative equilibrium to achieve this effect.  With a smaller adiabatic temperature gradient, our model develops two convective zones within the atmosphere. In the three-band model, vertical mixing creates a pseudo-adiabatic zone that compresses the lower convective layer and reduces the inversion area's size. But the convective flux will decrease the bottom temperature.
In the lower-Rosseland-opacity atmosphere, convective energy flux can expand the width of the outermost convective zone, indirectly reducing the temperature of the radiative layer.
While the lower atmosphere's temperature decreases slightly compared to cases with only vertical mixing, the PCB and PBR positions shift outward. Additionally, the upper atmospheric temperature drops significantly.

The increase in the visible band influences the temperature profile significantly. 
In the five-band picket fence model, the choice of the gamma parameters determines the combined effect of the vertical mixing and convection.
When gamma parameters remain constant, a steady increase in the temperature profile enables convection and mixing to pull the deeper convective layer inward, which results in cooling in the middle atmosphere.
Consequently, the convective layer in the upper atmosphere expands slightly. In low-Rosseland-opacity atmospheres, reduced opacity enhances the vertical mixing energy flux at the base of the atmosphere, which expands the pseudo-adiabatic region and pushes the innermost convective zone inward. 
Chemical elements can change multiple gamma parameters, which in turn affects the temperature profile and convective processes.
Studying the temperature profile both with and without TiO/VO shows how convection and vertical mixing cause changes, as illustrated in Figure~\ref{fig_Pseudo_rcb}.
However, keeping the full temperature profile in line with the chemical composition seems to limit the possible values for the vertical mixing coefficient $K_{\rm zz}$.

We propose that the incorporation of vertical mixing and convective energy has the potential to continuously modify the atmospheric structure and the chemical distribution of the planet, as chemical elements are particularly sensitive to variations in temperature profile. Moreover, the inclusion of additional chemical factors is expected to induce further alterations.

This study improves understanding of atmospheric energy flux equilibrium and lays groundwork for future research on atmospheric chemical distribution. Despite addressing chemical element effects, limitations exist. We consider their impact on parameters $\gamma$, as defined in \S~\ref{subsec_chemistry}, with constant $\kappa_{\rm R}$. Reactions involving carbon, hydrogen, oxygen, and nitrogen \citep{Heng2016ApJ,Gandhi2017}, and other processes, alter relative abundances, affecting opacity and temperature distribution. We use only five radiative transfer bands, while comprehensive models use up to 196 frequency bins \citep{Parmentier2015} for a two-stream model \citep{Toon1989JGR}, impacting radiative-convective equilibrium. 
To overcome these limitations, we'll develop a multi-band model with Genesis code \citet{Gandhi2017} in the future, featuring line-by-line opacities \citep{Freedman2008ApJS, Valencia2013ApJ} and thermochemical equilibrium \citep{Gandhi2017,2017MNRAS...Hubeny}, for precise vertical mixing and a firm basis for observational comparisons.

The interaction between thermochemical equilibrium and the RCME can also influence the formation of clouds \citep{WallaceOrmel2019A&A,Lee2024arXiv}, haze \citep{Wordsworth2022A}, and phase transitions. Consequently, this relationship affects the determination of the transition radius \citep{Hansen2008ApJS}, the prediction of secondary eclipses \citep{MacDonald2019PhDT}, and the characteristics of the transmission spectrum. Furthermore, convection, influenced by chemical interactions, can modify the properties of the convective layer, which is related to realistic moist convection \citep{Seeley2023PSJ}. However, moist convection may become unstable due to convective inhibition \citep{Cavali2017Icar,Leconte2024A&A}. Additionally, aerosol scattering impacts these atmospheric properties. Moreover, the RCME model can be utilized to investigate planets with H2-rich atmospheres, which are crucial for assessing their habitability \citep {Wordsworth2022A} .

\section{Acknowledgments}
     We thank the anonymous referee for the helpful comments and suggestions that clarified and improved the manuscript.
     Fruitful discussions with Jianheng Guo are highly appreciated. 
     This work has been supported by the National SKA Program of China (grant No. 2022SKA0120101), the National Natural Science Foundation of China (NSFC, No. 12288102), 
     and National Key R \& D Program of China (No. 2020YFC2201200) and the science research grants from the China Manned Space Project (No. CMS-CSST-2021-B09, CMS-CSST-2021-B12, and CMS-CSST-2021-A10), 
     International Centre of Super-novae,Yunnan Key Laboratory(No.202302AN360001), Guangdong Basic and Applied Basic Research Foundation (grant 2023A1515110805), and the grants from the opening fund of State Key Laboratory of Lunar and Planetary Sciences (Macau University of Science and Technology, Macau FDCT Grant No. SKL-LPS(MUST)-2021-2023). C.Y. has been supported by the National Natural Science Foundation of China (grants 11521303, 11733010, 11873103, and 12373071).
     Dong-Dong Ni has been supported by the  National Natural Science Foundation of China (grant No. 12022517), the Science and Technology Development Fund, Macau SAR (File No. 0139/2024/RIA2 ).
        

\bibliography{sample631}{}

\begin{thebibliography}{}
\expandafter\ifx\csname natexlab\endcsname\relax\def\natexlab#1{#1}\fi
\providecommand{\url}[1]{\href{#1}{#1}}
\providecommand{\dodoi}[1]{doi:~\href{http://doi.org/#1}{\nolinkurl{#1}}}
\providecommand{\doeprint}[1]{\href{http://ascl.net/#1}{\nolinkurl{http://ascl.net/#1}}}
\providecommand{\doarXiv}[1]{\href{https://arxiv.org/abs/#1}{\nolinkurl{https://arxiv.org/abs/#1}}}

\bibitem[{{Auer} \& {Mihalas}(1969)}]{1969ApJAuer...Mihalas}
{Auer}, L.~H., \& {Mihalas}, D. 1969, \apj, 158, 641, \dodoi{10.1086/150226}

\bibitem[{{Carter} {et~al.}(2023){Carter}, {Hinkley}, {Kammerer}, {Skemer},
  {Biller}, {Leisenring}, {Millar-Blanchaer}, {Petrus}, {Stone}, {Ward-Duong},
  {Wang}, {Girard}, {Hines}, {Perrin}, {Pueyo}, {Balmer}, {Bonavita},
  {Bonnefoy}, {Chauvin}, {Choquet}, {Christiaens}, {Danielski}, {Kennedy},
  {Matthews}, {Miles}, {Patapis}, {Ray}, {Rickman}, {Sallum}, {Stapelfeldt},
  {Whiteford}, {Zhou}, {Absil}, {Boccaletti}, {Booth}, {Bowler}, {Chen},
  {Currie}, {Fortney}, {Grady}, {Greebaum}, {Henning}, {Hoch}, {Janson},
  {Kalas}, {Kenworthy}, {Kervella}, {Kraus}, {Lagage}, {Liu}, {Macintosh},
  {Marino}, {Marley}, {Marois}, {Matthews}, {Mawet}, {McElwain}, {Metchev},
  {Meyer}, {Molliere}, {Moran}, {Morley}, {Mukherjee}, {Pantin}, {Quirrenbach},
  {Rebollido}, {Ren}, {Schneider}, {Vasist}, {Worthen}, {Wyatt},
  {Briesemeister}, {Bryan}, {Calissendorff}, {Cantalloube}, {Cugno}, {De
  Furio}, {Dupuy}, {Factor}, {Faherty}, {Fitzgerald}, {Franson}, {Gonzales},
  {Hood}, {Howe}, {Kuzuhara}, {Lagrange}, {Lawson}, {Lazzoni}, {Lew}, {Liu},
  {Llop-Sayson}, {Lloyd}, {Martinez}, {Mazoyer}, {Palma-Bifani}, {Quanz},
  {Redai}, {Samland}, {Schlieder}, {Tamura}, {Tan}, {Uyama}, {Vigan}, {Vos},
  {Wagner}, {Wolff}, {Ygouf}, {Zhang}, {Zhang}, \& {Zhang}}]{2023Carter}
{Carter}, A.~L., {Hinkley}, S., {Kammerer}, J., {et~al.} 2023, \apjl, 951, L20,
  \dodoi{10.3847/2041-8213/acd93e}

\bibitem[{{Cavali{\'e}} {et~al.}(2017){Cavali{\'e}}, {Venot}, {Selsis},
  {Hersant}, {Hartogh}, \& {Leconte}}]{Cavali2017Icar}
{Cavali{\'e}}, T., {Venot}, O., {Selsis}, F., {et~al.} 2017, \icarus, 291, 1,
  \dodoi{10.1016/j.icarus.2017.03.015}

\bibitem[{{Chandrasekhar}(1935)}]{Chandrasekhar1935}
{Chandrasekhar}, S. 1935, \mnras, 96, 21, \dodoi{10.1093/mnras/96.1.21}

\bibitem[{{Fortney} {et~al.}(2005){Fortney}, {Marley}, {Lodders}, {Saumon}, \&
  {Freedman}}]{Fortney2005}
{Fortney}, J.~J., {Marley}, M.~S., {Lodders}, K., {Saumon}, D., \& {Freedman},
  R. 2005, \apjl, 627, L69, \dodoi{10.1086/431952}

\bibitem[{{Freedman} {et~al.}(2008){Freedman}, {Marley}, \&
  {Lodders}}]{Freedman2008ApJS}
{Freedman}, R.~S., {Marley}, M.~S., \& {Lodders}, K. 2008, \apjs, 174, 504,
  \dodoi{10.1086/521793}

\bibitem[{{Fromang} {et~al.}(2016){Fromang}, {Leconte}, \&
  {Heng}}]{2016AA...Fromang}
{Fromang}, S., {Leconte}, J., \& {Heng}, K. 2016, \aap, 591, A144,
  \dodoi{10.1051/0004-6361/201527600}

\bibitem[{{Gandhi} \& {Madhusudhan}(2017)}]{Gandhi2017}
{Gandhi}, S., \& {Madhusudhan}, N. 2017, \mnras, 472, 2334,
  \dodoi{10.1093/mnras/stx1601}

\bibitem[{{Gandhi}(2019)}]{2019PhDT...Gandhi}
{Gandhi}, S.~N. 2019, PhD thesis, University of Cambridge, UK

\bibitem[{{Guillot}(2010)}]{2010A&A...Guillot}
{Guillot}, T. 2010, \aap, 520, A27, \dodoi{10.1051/0004-6361/200913396}

\bibitem[{{Gupta} {et~al.}(2022){Gupta}, {Nicholson}, \&
  {Schlichting}}]{2022MNRAS.Gupta}
{Gupta}, A., {Nicholson}, L., \& {Schlichting}, H.~E. 2022, \mnras, 516, 4585,
  \dodoi{10.1093/mnras/stac2488}

\bibitem[{{Hansen}(2008)}]{Hansen2008ApJS}
{Hansen}, B. M.~S. 2008, \apjs, 179, 484, \dodoi{10.1086/591964}

\bibitem[{{Heng} \& {Tsai}(2016)}]{Heng2016ApJ}
{Heng}, K., \& {Tsai}, S.-M. 2016, \apj, 829, 104,
  \dodoi{10.3847/0004-637X/829/2/104}

\bibitem[{{Holton}(1984)}]{1984maph...Holton}
{Holton}, J.~R. 1984, in Middle Atmosphere Program. Handbook for MAP. Volume
  14, Vol.~14, 359

\bibitem[{{Hubeny}(2017)}]{2017MNRAS...Hubeny}
{Hubeny}, I. 2017, \mnras, 469, 841, \dodoi{10.1093/mnras/stx758}

\bibitem[{{Hubeny} {et~al.}(2003){Hubeny}, {Burrows}, \&
  {Sudarsky}}]{Hubeny2003ApJ}
{Hubeny}, I., {Burrows}, A., \& {Sudarsky}, D. 2003, \apj, 594, 1011,
  \dodoi{10.1086/377080}

\bibitem[{{Hubeny} \& {Mihalas}(2014)}]{2014tsa..book...Hubeny}
{Hubeny}, I., \& {Mihalas}, D. 2014, {Theory of Stellar Atmospheres}

\bibitem[{{Kippenhahn} {et~al.}(2013){Kippenhahn}, {Weigert}, \&
  {Weiss}}]{2013book..Kippenhahn}
{Kippenhahn}, R., {Weigert}, A., \& {Weiss}, A. 2013, {Stellar Structure and
  Evolution}, \dodoi{10.1007/978-3-642-30304-3}

\bibitem[{{Knutson} {et~al.}(2010){Knutson}, {Howard}, \&
  {Isaacson}}]{Knutson2010ApJ}
{Knutson}, H.~A., {Howard}, A.~W., \& {Isaacson}, H. 2010, \apj, 720, 1569,
  \dodoi{10.1088/0004-637X/720/2/1569}

\bibitem[{{Leconte}(2018)}]{2018Leconte}
{Leconte}, J. 2018, \apjl, 853, L30, \dodoi{10.3847/2041-8213/aaaa61}

\bibitem[{{Leconte} {et~al.}(2024){Leconte}, {Spiga}, {Cl{\'e}ment}, {Guerlet},
  {Selsis}, {Milcareck}, {Cavali{\'e}}, {Moreno}, {Lellouch},
  {Carri{\'o}n-Gonz{\'a}lez}, {Charnay}, \& {Lef{\`e}vre}}]{Leconte2024A&A}
{Leconte}, J., {Spiga}, A., {Cl{\'e}ment}, N., {et~al.} 2024, \aap, 686, A131,
  \dodoi{10.1051/0004-6361/202348928}

\bibitem[{{Lee} \& {Ohno}(2024)}]{Lee2024arXiv}
{Lee}, E. K.~H., \& {Ohno}, K. 2024, arXiv e-prints, arXiv:2411.10305.
\newblock \doarXiv{2411.10305}

\bibitem[{{Lindzen}(1981)}]{Lindzen1981}
{Lindzen}, R.~S. 1981, \jgr, 86, 9707, \dodoi{10.1029/JC086iC10p09707}

\bibitem[{{Lodders}(2002)}]{Lodders2002ApJ}
{Lodders}, K. 2002, \apj, 577, 974, \dodoi{10.1086/342241}

\bibitem[{{MacDonald}(2019)}]{MacDonald2019PhDT}
{MacDonald}, R.~J. 2019, PhD thesis, University of Cambridge, UK

\bibitem[{{Madhusudhan} {et~al.}(2012){Madhusudhan}, {Lee}, \&
  {Mousis}}]{Madhusudhan2012ApJ}
{Madhusudhan}, N., {Lee}, K. K.~M., \& {Mousis}, O. 2012, \apjl, 759, L40,
  \dodoi{10.1088/2041-8205/759/2/L40}

\bibitem[{{Menou}(2019)}]{2019MNRAS...Menou}
{Menou}, K. 2019, \mnras, 485, L98, \dodoi{10.1093/mnrasl/slz041}

\bibitem[{{Mihalas} \& {Mihalas}(1984)}]{Mihalas..1984}
{Mihalas}, D., \& {Mihalas}, B.~W. 1984, {Foundations of radiation
  hydrodynamics}

\bibitem[{{Miles} {et~al.}(2023){Miles}, {Biller}, {Patapis}, {Worthen},
  {Rickman}, {Hoch}, {Skemer}, {Perrin}, {Whiteford}, {Chen}, {Sargent},
  {Mukherjee}, {Morley}, {Moran}, {Bonnefoy}, {Petrus}, {Carter}, {Choquet},
  {Hinkley}, {Ward-Duong}, {Leisenring}, {Millar-Blanchaer}, {Pueyo}, {Ray},
  {Sallum}, {Stapelfeldt}, {Stone}, {Wang}, {Absil}, {Balmer}, {Boccaletti},
  {Bonavita}, {Booth}, {Bowler}, {Chauvin}, {Christiaens}, {Currie},
  {Danielski}, {Fortney}, {Girard}, {Grady}, {Greenbaum}, {Henning}, {Hines},
  {Janson}, {Kalas}, {Kammerer}, {Kennedy}, {Kenworthy}, {Kervella}, {Lagage},
  {Lew}, {Liu}, {Macintosh}, {Marino}, {Marley}, {Marois}, {Matthews},
  {Matthews}, {Mawet}, {McElwain}, {Metchev}, {Meyer}, {Molliere}, {Pantin},
  {Quirrenbach}, {Rebollido}, {Ren}, {Schneider}, {Vasist}, {Wyatt}, {Zhou},
  {Briesemeister}, {Bryan}, {Calissendorff}, {Cantalloube}, {Cugno}, {De
  Furio}, {Dupuy}, {Factor}, {Faherty}, {Fitzgerald}, {Franson}, {Gonzales},
  {Hood}, {Howe}, {Kraus}, {Kuzuhara}, {Lagrange}, {Lawson}, {Lazzoni}, {Liu},
  {Llop-Sayson}, {Lloyd}, {Martinez}, {Mazoyer}, {Quanz}, {Redai}, {Samland},
  {Schlieder}, {Tamura}, {Tan}, {Uyama}, {Vigan}, {Vos}, {Wagner}, {Wolff},
  {Ygouf}, {Zhang}, {Zhang}, \& {Zhang}}]{2023Miles}
{Miles}, B.~E., {Biller}, B.~A., {Patapis}, P., {et~al.} 2023, \apjl, 946, L6,
  \dodoi{10.3847/2041-8213/acb04a}

\bibitem[{{Ormel} \& {Min}(2019)}]{WallaceOrmel2019A&A}
{Ormel}, C.~W., \& {Min}, M. 2019, \aap, 622, A121,
  \dodoi{10.1051/0004-6361/201833678}

\bibitem[{{Parmentier}(2014)}]{Parmentier2014PhDT}
{Parmentier}, V. 2014, PhD thesis, Laboratoire Universitaire d'Astrophysique de
  Nice; Observatoire de la Cote d'Azur, France

\bibitem[{{Parmentier} \& {Guillot}(2014)}]{2014A&A...Parmentier}
{Parmentier}, V., \& {Guillot}, T. 2014, \aap, 562, A133,
  \dodoi{10.1051/0004-6361/201322342}

\bibitem[{{Parmentier} {et~al.}(2015){Parmentier}, {Guillot}, {Fortney}, \&
  {Marley}}]{Parmentier2015}
{Parmentier}, V., {Guillot}, T., {Fortney}, J.~J., \& {Marley}, M.~S. 2015,
  \aap, 574, A35, \dodoi{10.1051/0004-6361/201323127}

\bibitem[{{Parmentier} {et~al.}(2013){Parmentier}, {Showman}, \&
  {Lian}}]{Parmentier2013}
{Parmentier}, V., {Showman}, A.~P., \& {Lian}, Y. 2013, \aap, 558, A91,
  \dodoi{10.1051/0004-6361/201321132}

\bibitem[{{Pelletier} {et~al.}(2023){Pelletier}, {Benneke}, {Ali-Dib},
  {Prinoth}, {Kasper}, {Seifahrt}, {Bean}, {Debras}, {Klein}, {Bazinet},
  {Hoeijmakers}, {Kesseli}, {Lim}, {Carmona}, {Pino}, {Casasayas-Barris},
  {Hood}, \& {St{\"u}rmer}}]{2023Natur.Pelletier}
{Pelletier}, S., {Benneke}, B., {Ali-Dib}, M., {et~al.} 2023, \nat, 619, 491,
  \dodoi{10.1038/s41586-023-06134-0}

\bibitem[{{Powell} \& {Zhang}(2024)}]{2024arXivZhang}
{Powell}, D., \& {Zhang}, X. 2024, arXiv e-prints, arXiv:2404.08759,
  \dodoi{10.48550/arXiv.2404.08759}

\bibitem[{{Robinson} \& {Catling}(2012)}]{Robinson2012Ap}
{Robinson}, T.~D., \& {Catling}, D.~C. 2012, \apj, 757, 104,
  \dodoi{10.1088/0004-637X/757/1/104}

\bibitem[{{Ryu} {et~al.}(2018){Ryu}, {Zingale}, \& {Perna}}]{2018MNRAS...Ryu}
{Ryu}, T., {Zingale}, M., \& {Perna}, R. 2018, \mnras, 481, 5517,
  \dodoi{10.1093/mnras/sty2638}

\bibitem[{{Saumon} {et~al.}(1995){Saumon}, {Hubbard}, {Burrows}, {Guillot}, \&
  {Lunine}}]{Saumon1995}
{Saumon}, D., {Hubbard}, W.~B., {Burrows}, A., {Guillot}, T., \& {Lunine},
  J.~I. 1995, in American Astronomical Society Meeting Abstracts, Vol. 187,
  American Astronomical Society Meeting Abstracts, 70.02

\bibitem[{{Seager}(2010)}]{2010Seager}
{Seager}, S. 2010, {Exoplanet Atmospheres: Physical Processes}

\bibitem[{{Seeley} \& {Wordsworth}(2023)}]{Seeley2023PSJ}
{Seeley}, J.~T., \& {Wordsworth}, R.~D. 2023, \psj, 4, 34,
  \dodoi{10.3847/PSJ/acb0cb}

\bibitem[{{Sing} {et~al.}(2024{\natexlab{a}}){Sing}, {Rustamkulov},
  {Thorngren}, {Barstow}, {Tremblin}, {Alves de Oliveira}, {Beck}, {Birkmann},
  {Challener}, {Crouzet}, {Espinoza}, {Ferruit}, {Giardino}, {Gressier}, {Lee},
  {Lewis}, {Maiolino}, {Manjavacas}, {Rauscher}, {Sirianni}, \&
  {Valenti}}]{2024Natur.630..831S}
{Sing}, D.~K., {Rustamkulov}, Z., {Thorngren}, D.~P., {et~al.}
  2024{\natexlab{a}}, \nat, 630, 831, \dodoi{10.1038/s41586-024-07395-z}

\bibitem[{{Sing} {et~al.}(2024{\natexlab{b}}){Sing}, {Rustamkulov},
  {Thorngren}, {Barstow}, {Tremblin}, {Alves de Oliveira}, {Beck}, {Birkmann},
  {Challener}, {Crouzet}, {Espinoza}, {Ferruit}, {Giardino}, {Gressier}, {Lee},
  {Lewis}, {Maiolino}, {Manjavacas}, {Rauscher}, {Sirianni}, \&
  {Valenti}}]{Sing2024Natur}
---. 2024{\natexlab{b}}, \nat, 630, 831, \dodoi{10.1038/s41586-024-07395-z}

\bibitem[{{Spiegel} {et~al.}(2009){Spiegel}, {Silverio}, \&
  {Burrows}}]{Spiegel2009}
{Spiegel}, D.~S., {Silverio}, K., \& {Burrows}, A. 2009, \apj, 699, 1487,
  \dodoi{10.1088/0004-637X/699/2/1487}

\bibitem[{{Strobel} {et~al.}(1987){Strobel}, {Summers}, {Bevilacqua}, {Deland},
  \& {Allen}}]{Strobel1987}
{Strobel}, D.~F., {Summers}, M.~E., {Bevilacqua}, R.~M., {Deland}, M.~T., \&
  {Allen}, M. 1987, \jgr, 92, 6691, \dodoi{10.1029/JD092iD06p06691}

\bibitem[{{Toon} {et~al.}(1989){Toon}, {McKay}, {Ackerman}, \&
  {Santhanam}}]{Toon1989JGR}
{Toon}, O.~B., {McKay}, C.~P., {Ackerman}, T.~P., \& {Santhanam}, K. 1989,
  \jgr, 94, 16287, \dodoi{10.1029/JD094iD13p16287}

\bibitem[{{Valencia} {et~al.}(2013){Valencia}, {Guillot}, {Parmentier}, \&
  {Freedman}}]{Valencia2013ApJ}
{Valencia}, D., {Guillot}, T., {Parmentier}, V., \& {Freedman}, R.~S. 2013,
  \apj, 775, 10, \dodoi{10.1088/0004-637X/775/1/10}

\bibitem[{{Wallace} \& {Hobbs}(1977)}]{Wallace1977}
{Wallace}, J.~M., \& {Hobbs}, P.~V. 1977, {Atmosphere science - an introductory
  survey.}

\bibitem[{{Welbanks} {et~al.}(2024){Welbanks}, {Bell}, {Beatty}, {Line},
  {Ohno}, {Fortney}, {Schlawin}, {Greene}, {Rauscher}, {McGill}, {Murphy},
  {Parmentier}, {Tang}, {Edelman}, {Mukherjee}, {Wiser}, {Lagage}, {Dyrek}, \&
  {Arnold}}]{Welbanks2024}
{Welbanks}, L., {Bell}, T.~J., {Beatty}, T.~G., {et~al.} 2024, \nat, 630, 836,
  \dodoi{10.1038/s41586-024-07514-w}

\bibitem[{{Wordsworth} \& {Kreidberg}(2022)}]{Wordsworth2022A}
{Wordsworth}, R., \& {Kreidberg}, L. 2022, \araa, 60, 159,
  \dodoi{10.1146/annurev-astro-052920-125632}

\bibitem[{{Youdin} \& {Mitchell}(2010)}]{Youdin2010}
{Youdin}, A.~N., \& {Mitchell}, J.~L. 2010, \apj, 721, 1113,
  \dodoi{10.1088/0004-637X/721/2/1113}

\bibitem[{Zhang(2020)}]{Zhang_2020}
Zhang, X. 2020, Research in Astronomy and Astrophysics, 20, 099,
  \dodoi{10.1088/1674-4527/20/7/99}

\bibitem[{{Zhang}(2023)}]{ZhangXi2023ApJ}
{Zhang}, X. 2023, \apj, 957, 20, \dodoi{10.3847/1538-4357/acee66}

\bibitem[{{Zhang} \& {Showman}(2018{\natexlab{a}})}]{2018ApJ...Zhang....1Z}
{Zhang}, X., \& {Showman}, A.~P. 2018{\natexlab{a}}, \apj, 866, 1,
  \dodoi{10.3847/1538-4357/aada85}

\bibitem[{{Zhang} \& {Showman}(2018{\natexlab{b}})}]{2018ApJ...Zhang}
---. 2018{\natexlab{b}}, \apj, 866, 2, \dodoi{10.3847/1538-4357/aada7c}

\bibitem[{{Zhong} {et~al.}(2025){Zhong}, {Zhang}, {Zhong}, {Ma}, {Tan}, \&
  {Yu}}]{Zhong2024}
{Zhong}, W., {Zhang}, Z.-T., {Zhong}, H.-S., {et~al.} 2025, \apj, 978, 4,
  \dodoi{10.3847/1538-4357/ad9473}

\end{thebibliography}
\bibliographystyle{aasjournal}

\end{document}